\documentclass[aip,reprint]{revtex4-1}

\usepackage{graphicx}
\usepackage{amssymb,amsmath,amsopn,amsfonts}
\usepackage{bbm}
\usepackage{physics}
\usepackage{hyperref}
\usepackage{xcolor}

\begin{document}


\title{Non-Hermiticity in quantum nonlinear optics through symplectic transformations}



\author{Ross~Wakefield}
\affiliation{Duality Quantum Photonics, 6 Lower Park Row, Bristol, BS1 5BJ, UK}
\affiliation{Quantum Engineering Technology Labs, H. H. Wills Physics Laboratory and Department of Electrical and Electronic Engineering, University of Bristol, Bristol BS8 1FD, UK}
\affiliation{Quantum Engineering Centre for Doctoral Training, H. H. Wills Physics Laboratory and Department of Electrical and Electronic Engineering, University of Bristol, BS8 1FD, UK}
\author{Anthony~Laing}
\affiliation{Duality Quantum Photonics, 6 Lower Park Row, Bristol, BS1 5BJ, UK}
\affiliation{Quantum Engineering Technology Labs, H. H. Wills Physics Laboratory and Department of Electrical and Electronic Engineering, University of Bristol, Bristol BS8 1FD, UK}
\author{Yogesh N. Joglekar}
\email{yojoglek@iu.edu}
\affiliation{Department of Physics, Indiana University Indianapolis (IUI), Indianapolis, Indiana 46202, USA}


\date{\today}

\begin{abstract}
    Over the past decade classical optical systems with gain or loss, modelled by non-Hermitian parity-time symmetric Hamiltonians, have been deeply investigated. Yet, their applicability to the quantum domain with number-resolved photonic states is fundamentally voided by quantum-limited amplifier noise. Here, we show that second-quantised Hermitian Hamiltonians on the Fock space give rise to non-Hermitian effective Hamiltonians that generate the dynamics of corresponding creation and annihilation operators. Using this equivalence between $\mathcal{PT}$-symmetry and symplectic Bogoliubov transformations, we create a quantum optical scheme comprising squeezing, phase-shifters, and beam-splitters for simulating arbitrary non-unitary processes by way of singular value decomposition. In contrast to the post-selection scheme for non-Hermitian quantum simulation, the success probability in this approach is independent of the system size or simulation time, and can be efficiently Trotterised similar to a unitary transformation.
\end{abstract}

\maketitle


\section{Introduction}
\label{sec:intro}

Since the seminal discovery by Bender and Boettcher 25 years ago~\cite{Bender1998}, the field of non-Hermitian Hamiltonians has flourished far and wide beyond its mathematical-physics niche~\cite{Bender2001,Mostafazadeh2002,Mostafazadeh2003-equivalence}. Among such Hamiltonians with generically complex eigenvalues, those invariant under combined operations of parity and time-reversal are called parity-time ($\mathcal{PT}$) symmetric Hamiltonians. Due to this antilinear symmetry, the eigenvalues of a $\mathcal{PT}$-symmetric, non-Hermitian Hamiltonian are real or complex conjugates, while the corresponding (right) eigenvectors are not orthogonal under the standard, Dirac inner-product. The transition from real to complex-conjugates spectrum, called $\mathcal{PT}$-symmetry breaking transition~\cite{Feng2017,Ozdemir2019}, occurs at an exceptional point (EP) degeneracy where the corresponding eigenvectors also coalesce~\cite{Kato1995,Miri2019}. Over the past decade, a Cambrian explosion of experimental platforms -- classical~\cite{Regensburger2012,Hodaei2014,Peng2014,Peng2014b,Feng2014,Zhu2014,Schindler2011,Wang2020,Quiroz2022} and quantum~\cite{Li2019,Wu2019,Klauck2019,Naghiloo2019,Ding2021} -- has made it clear that $\mathcal{PT}$-symmetric Hamiltonians accurately model open systems with balanced, separated gain and loss. 

Balanced gain and losses, easy to implement in classical systems, face challenges in the quantum domain. Even at zero temperature, where thermal fluctuations accompanying the dissipation~\cite{Kubo1966} are eliminated, the gain process is accompanied by the inevitable quantum noise~\cite{Caves1982,Scheel2018}. Several approaches have been used to circumvent this fundamental problem and thereby simulate $\mathcal{PT}$-symmetric Hamiltonians in the quantum domain. They include using mode-selective losses~\cite{guo09,Klauck2019}, embedding the non-Hermitian Hamiltonian into a larger Hermitian Hamiltonian with an ancilla~\cite{Gunther2008-proposal,Wu2019}, dilating the non-unitary time evolution operator into a larger unitary~\cite{halmos-1950,Nicola2022}, or post-selecting on no-quantum-jump trajectories in small, dissipative quantum systems~\cite{Naghiloo2019,Ding2021,Quinn2023}. The dynamics of such systems is governed by norm-preserving, nonlinear Schr\"{o}dinger equation $i\partial_t\ket{\psi(t)}=(H_0+i\Gamma-i\bra{\psi}\Gamma\ket{\psi})\ket{\psi(t)}$ where $H_0$ and $\Gamma$ denote the Hermitian and anti-Hermitian parts of the Hamiltonian $H_0+i\Gamma$ respectively~\cite{Brody2012,Varma2023}. Appropriating salient features of unitary and Lindbladian dynamics, the coherent, non-unitary evolution generated by a non-Hermitian Hamiltonian keeps pure states pure, but changes the entropy of mixed states~\cite{Xue2020}. Although post-selection allows one to decipher non-Hermitian effects in the transient dynamics of an open, dissipative system, its applicability to large system sizes or long simulation times is severely restricted by an exponentially vanishing set of quantum trajectories with no quantum jump.

While a small quantum system is described by an evolving state $\ket{\psi(t)}$ (or a density matrix), a bosonic system in the quantum domain is characterised by its annihilation and creation operators $a_j,a^\dagger_k$ for modes $j$ and $k$ respectively, that satisfy the canonical commutation relations $[a_j,a^{\dagger}_k]=\delta_{jk}$. The equations of motion for these operators are generated by the second-quantized, Hermitian Hamiltonian for the system. Here, we connect the unitary dynamical transformation on the Hilbert state-space $\mathcal H$ (Fock space) with a transformation on the annihilation, creation operators that define an operator space $\mathcal{S}\subset \mathcal{B}\left(\mathcal{H}\right)$. For a single photon propagating in a linear interferometer, implemented with a Reck~\cite{Reck1994Experimental} or Clements \cite{Clements2016Optimal} scheme, the state-transformation on the Fock space $\mathcal{H}$ and the operator transformation on $\mathcal{S}\subset \mathcal{B}\left(\mathcal{H}\right)$ are both unitary. However, this is not always true of nonlinear-optics transformations, where the transformation on the Hilbert space $\mathcal{H}$ is unitary while the transformation on the operator space $\mathcal{S}$ is instead symplectic and non-unitary~\cite{Arvind1995,Wnsche2000,Braunstein2005Squeezing}. By characterising the non-Hermitian effective Hamiltonian $H_\mathrm{eff}$ responsible for it, we show that key features of $\mathcal{PT}$-symmetric Hamiltonians such as the EP degeneracy can be simulated with quantum nonlinear optics. Coupling to ancilla modes allows for recreating any non-unitary transformation, and we give a method to simulate these transformations using squeezing and linear interferometers. This simulation method can be Trotterised to give an efficient method for simulating coherent, non-unitary dynamics of bosonic modes in the quantum domain. 

The plan of the paper is as follows. In Sec.~\ref{sec:tms} we present an example with two coupled bosonic modes interacting via Hermitian Hamiltonians. Section~\ref{sec:general} describes the general formalism for $N$ interacting modes, the structure of the resulting effective, non-Hermitian matrix $H_\mathrm{eff}$, and its antilinear symmetries. Results for time-dependent occupation numbers are also presented. In Sec.~\ref{sec:simulation}, we present the scheme for simulating arbitrary non-Hermitian evolution over $N$ modes by using singular value decomposition (SVD). The paper is concluded with a brief discussion in Sec.~\ref{sec:disc}.


\section{Non-Hermiticity in a two-mode system}
\label{sec:tms}
Let us consider a system of two coupled bosonic modes governed by Hamiltonians ($\hbar=1$)
\begin{align}
\label{eq:bs}
&H_\mathrm{bs}=\omega_1 a^{\dagger}_1 a_1+\omega_2 a^{\dagger}_2 a_2+ g(a^\dagger_1 a_2+a^{\dagger}_2 a_1)=H^\dagger_{\mathrm{bs}}\\
\label{eq:tms}
&H_\mathrm{tms}=\omega_1 a^{\dagger}_1 a_1+\omega_2 a^{\dagger}_2 a_2 +i\kappa (a_1 a_2- a^{\dagger}_2a^{\dagger}_1)=H^\dagger_{\mathrm{tms}}
\end{align}
where $\omega_1,\omega_2$ are their fundamental frequencies, $g\in\mathbb{R}$ denotes the amplitude of mixing between the two modes, and $\kappa\in\mathbb{R}$ is the strength of two-mode squeezing process enabled by an external, classical pump. Although both Hamiltonians are Hermitian, the total boson-number operator $a^\dagger_1a_1+a^\dagger_2a_2$ commutes with only the "beam-splitter" Hamiltonian \eqref{eq:bs}. For the rest of the article, we will use the notation $\ket{A(t)}=[a^\dagger_1,a^\dagger_2,a_1,a_2]^T$ for a column vector with elements from the operator space. We will also use the equation of motion for an arbitrary operator $\Box$ to define the effective Hamiltonian on the operator space
\begin{align}
    \label{eq:heffdef}
    i\partial_t\Box=[H,\Box]\equiv H_\mathrm{eff}\Box.    
\end{align}
Thus, the superoperator $H_\mathrm{eff}$ encodes the adjoint action by the Hamiltonian $H$. The dynamics of mode operators in a system the beam-splitter Hamiltonian \eqref{eq:bs} becomes
\begin{align}
\label{eq:heffbs}
    i\partial_t\left[\begin{array}{c}
    a^{\dagger}_1\\
    a^{\dagger}_2\\
    a_1\\
    a_2
    \end{array}\right]=\left[\begin{array}{cccc}
       \omega_1  & g & 0 & 0  \\
       g &\omega_2 & 0 & 0\\
       0 & 0 & -\omega_1 & -g\\
       0 & 0 & -g & -\omega_2
    \end{array}\right]\left[\begin{array}{c}
    a^{\dagger}_1\\
    a^{\dagger}_2\\
    a_1\\
    a_2
    \end{array}\right].
\end{align}
It can be compactly written as $i\partial_t\ket{A(t)}=H_\mathrm{eff,bs}\ket{A(t)}$ with 
\begin{align}
\label{eq;heffbs}
    H_\mathrm{eff,bs}=\sigma_z\otimes\left[\omega_0\mathbbm{1}_2+\Delta\omega\sigma_z+g\sigma_x\right]=H^\dagger_{\mathrm{eff,bs}}
\end{align}
where $\omega_0=(\omega_1+\omega_2)/2$ is the average mode frequency and $\Delta\omega=(\omega_1-\omega_2)/2$ is the mode detuning. The four real  eigenvalues $\lambda_k=\pm(\omega_0\pm\sqrt{\Delta\omega^2+g^2})$ of $H_\mathrm{eff,bs}$ become doubly-degenerate at $g_\mathrm{DP}=\sqrt{\omega_0^2-\Delta\omega^2}$, but this is a diabolic-point degeneracy where the eigenvectors of the $4\times 4$ matrix $H_\mathrm{eff,bs}$ remain orthogonal. Thus, the transformations $M_\mathrm{bs}(t)=\exp(-iH_\mathrm{eff,bs}t)$ induced by $H_\mathrm{eff,bs}$ are also unitary. 

For a two-mode system with squeezing, Eq.\eqref{eq:tms}, the corresponding effective Hamiltonian is given by
\begin{align}
\label{eq:hefftms}
    H_\mathrm{eff,tms}=\sigma_z\otimes[\omega_0\mathbbm{1}_2+\Delta\omega\sigma_z]+i\kappa\sigma_x\otimes\sigma_x.
\end{align}
The Hamiltonian \eqref{eq:hefftms} is evidently not Hermitian, but it is invariant under combined actions of the linear parity operator $\mathcal{P}=\sigma_z\otimes\mathbbm{1}_2=\mathcal{P}^{-1}=\mathcal{P}^\dagger$ and the antilinear time-reversal operator $\mathcal{T}=*$ (complex conjugation). Its eigenvalues are given by $\lambda_k=\pm\Delta\omega\pm\sqrt{\omega_0^2-\kappa^2}$. Purely real at small pump-strengths $\kappa<\omega_0$, they become doubly-degenerate at $\kappa_\mathrm{EP}=\omega_0$, and then complex-conjugates as $\kappa$ exceeds the threshold $\kappa_\mathrm{EP}$. At this second-order exceptional-point (EP) degeneracy, due to coalescence of eigenvectors, the rank of $H_\mathrm{eff,tms}$ reduces from 4 to 2. Thus, the transformations $M_\mathrm{tms}(t)=\exp(-iH_\mathrm{eff,tms}t)$ induced by the Hermitian Hamiltonian $H_\mathrm{eff,tms}$ on the operator space are not unitary. As we will show in the following section, this simple example provides a general recipe to simulate non-Hermitian quantum dynamics by using Hermitian, non-linear (three-wave or four-wave mixing) processes in bosonic systems. 


\section{PT-symmetry and Symplectic Transformations in an N-mode system}
\label{sec:general}

The general bilinear, Hermitian Hamiltonian for $N$ bosonic modes is a sum of four terms, each representing a distinct physical process,
\begin{align}
\label{eq:H1}
\nonumber
    H=&\sum_{p=1}^N\omega_p a^\dagger_p a_p+\sum_{p\neq q=1}^N g_{pq}a^{\dagger}_p a_q\\
    +& \frac{1}{2}\sum_{p=1}^N\left(\kappa_p a^2_p+\mathrm{h.c.}\right)+\sum_{p\neq q=1}^N\left(\kappa_{pq} a_p a_q +\mathrm{h.c.}\right). 
\end{align}
The first term denotes a free phase-evolution Hamiltonian; the second term, which reciprocally swaps modes $p\leftrightarrow q$ ($g_{pq}=g^*_{qp}$) is a beam-splitter Hamiltonian; the third-term, annihilating or creating a pair of identical bosons, represents single-mode squeezing; and the last term corresponds to two-mode squeezing. Ignoring a zero-point energy term, the Hamiltonian \eqref{eq:H1} can be compactly written as $\bra{A}H_{2N}\ket{A}/2$ where $\bra{A}=[a_j,a^\dagger_j]$ is the operator row-vector and $H_{2N}$ is a $2N$-dimensional Hermitian matrix. The equation of motion for the $2N$-dimensional column vector $\ket{A}$ is given by $i\partial_t\ket{A(t)}=H_\mathrm{eff}\ket{A(t)}$ with 
\begin{align}
\label{eq:Heff}
\nonumber
   H_\mathrm{eff}&=\left[\begin{array}{cc}
      \mathbb{W}^T  & \mathbb{K} \\
      -\mathbb{K}^\dagger & -\mathbb{W}
   \end{array}\right]=\mathbbm{1}_2\otimes\frac{(\mathbb{W}^T-\mathbb{W})}{2}\\
   &+\sigma_z\otimes\frac{(\mathbb{W}^T+\mathbb{W})}{2}+i\sigma_y\otimes\mathbb{K}_H+i\sigma_x\otimes\mathbb{K}_{AH},
\end{align}
where $\mathbb{W}_{pq}=\omega_p\delta_{pq}+g_{pq}(1-\delta_{pq})$ is a Hermitian matrix, and $\mathbb{K}_{pq}=\kappa_p\delta_{pq}+\kappa_{pq}(1-\delta_{pq})=\mathbb{K}^T_{pq}$ is a complex, symmetric matrix with $\mathbb{K}_H=(\mathbb{K}+\mathbb{K}^\dagger)/2$ and $i\mathbb{K}_{AH}=(\mathbb{K}-\mathbb{K}^\dagger)/2$ as its Hermitian and anti-Hermitian parts respectively. It is clear that $H_\mathrm{eff}$ is Hermitian if and only if the bilinear Hamiltonian \eqref{eq:H1} has no single-mode or two-mode squeezing terms. On the other hand, $H_\mathrm{eff}$ is anti-Hermitian --- representing pure amplification or absorption processes --- if and only if there are no beam-splitter or phase-shifter terms. We also note that the bosonic number operator $\hat{N}(t)=\sum_{p=1}^N a^\dagger_p a_p=\bra{A(t)}\ket{A(t)}/2-N$ remains invariant if and only if it commutes with the Hamiltonian \eqref{eq:H1} or equivalently $\mathbb{K}=0$. 

To characterise the symmetries of $H_\mathrm{eff}$, Eq.\eqref{eq:Heff}, we consider the simplest case with $\mathbb{K}_H=0$ and $\mathbb{W}^T=\mathbb{W}$. (It is straightforward, albeit tedious, to carry out the analysis when all four terms in \eqref{eq:Heff} are nonzero.) The Hamiltonian $H_\mathrm{eff}=\sigma_z\otimes\mathbb{W}+i\sigma_x\otimes K_{AH}$ anticommutes with the Hermitian operator $\Pi=\sigma_y\otimes\mathbbm{1}_N=\Pi^{-1}$, and commutes with the antilinear operator $\mathcal{PT}=(\sigma_z\otimes\mathbbm{1}_N)*$. Due to the presence of both Chiral and antilinear symmetries, the eigenvalues of $H_\mathrm{eff}$ are particle-hole symmetric and complex-conjugates~\cite{Joglekar2010a}, i.e. if $\lambda$ is an eigenvalue of $H_\mathrm{eff}$ so are $-\lambda,\lambda^*$. Thus, a $\mathcal{PT}$-symmetric generator of time-evolution on the space $\mathcal{S}$ naturally arises from a Hermitian, second-quantized Hamiltonian in the presence of squeezing. 

The non-unitary time evolution operator $M(t)\equiv\exp(-iH_\mathrm{eff}t)$ satisfies $\Pi M\Pi=M^{-1}$. Although not unitary, it preserves the canonical commutation relations between the time-evolved creation and annihilation operators, $[a^\dagger_j(t),a_k(t)]=\delta_{jk}$ and $[a^\dagger_j(t),a^\dagger_k(t)]=0$. This constraint implies that $M(t)=M^T(t)$ satisfies
\begin{align}
    \label{eq:symp}
    M(t)\Omega M^T(t)=\Omega
\end{align}
with the antisymmetric form $\Omega=i\sigma_y\otimes\mathbbm{1}_N$, i.e. $M(t)$ is a complex, symplectic matrix~\cite{Arvind1995,Wnsche2000,Braunstein2005Squeezing} with unit determinant. The symplectic nature of maps that preserve bosonic canonical commutation relation, including unitary maps and bosonic Bogolioubov transformations, have been extensively investigated in the literature~\cite{Balian1969,Colpa1978,Tsallis1978,Broadbridge1981}. Interpreting them in terms of a Hamiltonian $H_\mathrm{eff}$ shows that starting from its Hermitian limit $\sigma_z\otimes\mathbb{W}$ with real eigenvalues, as the squeezing term $\mathbb{K}$ is increased, the spectrum of the effective Hamiltonian transitions from purely real ($\mathcal{PT}$-symmetric phase) to complex conjugates ($\mathcal{PT}$-broken phase). 

As another explicit example, we consider a two-mode system with continuous coupling between modes $g_{21}=-g_{12}=i\omega_0$ and simultaneous single-mode squeezing of both modes in opposite directions, $\kappa_1=i\kappa=-\kappa_2$. The effective Hamiltonian is now given by 
\begin{align}
\label{eq:hs}
H_\mathrm{eff,sms}&=i\left[\begin{array}{cccc}
0 & \omega_0 & \kappa & 0 \\ -\omega_0 & 0 & 0 & -\kappa \\ \kappa & 0 & 0 & \omega_0 \\ 0 & -\kappa & -\omega_0 & 0
\end{array}\right] \nonumber\\
&=\omega_0\mathbbm{1}_2\otimes\sigma_y+i\kappa\sigma_x\otimes\sigma_z
\end{align}
which has doubly-degenerate eigenvalues $\pm\sqrt{\omega_0^2-\kappa^2}=\pm i\Delta$ that transition from real to purely imaginary when the squeezing strength $\kappa$ exceeds the "beam-splitter" strength $\omega_0$. Hamiltonian \eqref{eq:hs} anticommutes with the operator $\Pi=\mathbbm{1}_2\otimes\sigma_x$ (Chiral symmetry), commutes with the operator $\sigma_x\otimes\mathbbm{1}_2$ (doubly degenerate eigenvalues), and is invariant under antilinear symmetries given by $\mathcal{P}_1\mathcal{T}=(\sigma_x\otimes\sigma_x)*$, $\mathcal{P}_2\mathcal{T}=(\mathbbm{1}_2\otimes\sigma_x)*$, and $\mathcal{P}_3\mathcal{T}=(\sigma_z\otimes\sigma_z)*$. In addition to its multiple $\mathcal{PT}$-symmetries, the Hamiltonian $H_\mathrm{eff,sms}$ also has an anti-$\mathcal{PT}$ symmetry; the operator $\mathcal{P}_4\mathcal{T}=(\sigma_x\otimes\mathbbm{1}_2)*$ {\it anticommutes} with the Hamiltonian $H_\mathrm{eff,sms}$. In general, when a Hamiltonian's spectrum changes from purely real ($\lambda_k\in\mathbb{R}$) to purely imaginary ($\lambda_k\in i\mathbb{R}$) across the EP, the Hamiltonian has both $\mathcal{PT}$ and anti-$\mathcal{PT}$ symmetries with distinct, suitably defined $\mathcal{P}$-operators~\cite{Ruzicka2021}. Note that, on the other hand, the two-mode squeezing Hamiltonian, Eq.\eqref{eq:hefftms}, does not have anti-$\mathcal{PT}$ symmetry (except when $\Delta\omega=0$ and the modes become degenerate). 

It is straightforward to obtain the non-unitary, symplectic time-evolution operator
\begin{align}
\label{eq:M}
    M(t)&=\mathbbm{1}_4\cosh(\Delta t) -iH_\mathrm{eff,sms}\frac{\sinh(\Delta t)}{\Delta}
\end{align}
that, at the second-order EP simplifies to the expected linear-in-time behaviour $M=\mathbbm{1}_4-itH_\mathrm{eff,sms}$. This explicit result allows calculation of time-dependent mode-occupations and variances. Starting from the vacuum state $\ket{0,0}$, we get 
\begin{align}
    \label{eq:n1n2}
    \langle N_1(t)\rangle &=\frac{\kappa^2}{\Delta^2}\sinh^2(\Delta t)=\langle N_2(t)\rangle,\\
    \nonumber
    (\Delta N_1)^2&=(\Delta N_2)^2\\
    \label{eq:varn1n2}
    &=\frac{\kappa^2}{\Delta^2}\sinh^2(\Delta t)\left[2\cosh^2(\Delta t)+\frac{\omega_0^2}{\Delta^2}\sinh^2(\Delta t)\right].
\end{align}
Across the EP, the system goes from a pure squeezer ($\kappa>\omega_0$) with monotonically increasing mode-occupations to a squeezed beam-splitter ($\kappa<\omega_0$) that shows oscillations in boson numbers. At the EP, the occupation numbers $\langle N_{1,2}(t)\rangle=(\kappa t)^2$ are much smaller than the fluctuations $\Delta N_{1,2}(t)=\sqrt{2}\kappa t$ at short times; at long times $\kappa t\gg1$, they both grow quadratically. These qualitative features hold for all bosonic quadratic Hamiltonians, Eq.\eqref{eq:H1}. 


\begin{figure*}
    \centering
	 \includegraphics[width=0.95\textwidth]{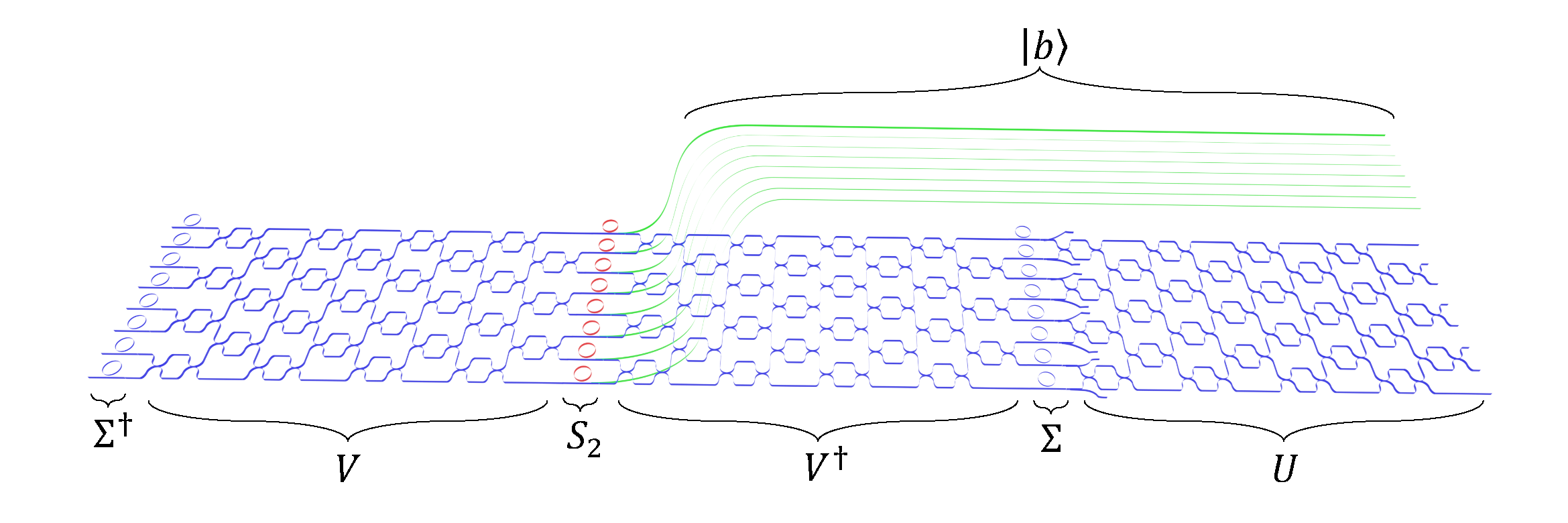}
    \caption{Diagram of an optical circuit implementing the singular-value-decomposition (SVD) of a symplectic transformation $M=M_UM_\Sigma M_{V^\dagger}$, Eq.\eqref{eq:symplectic}, over $N=8$ modes. $\Sigma^\dagger$ is a single-mode squeezer; $V$ is a linear optical circuit in Clements scheme~\cite{Clements2016Optimal}; $S_2$, a two-mode squeezer (red) with heralded $\ket{b}$-modes (green), acts as a creation operator $a^\dagger_j$; $V^\dagger$ is a linear optical circuit, followed by single-mode squeezer $\Sigma$ and linear optical circuit $U$.}
    \label{fig:diagram}
\end{figure*}

\section{Simulating A Non-Hermitian evolution}
\label{sec:simulation}

The standard quantum simulation methods~\cite{Buluta2009,Altman2021,Daley2022} can be divided into two classes. The first comprises continuous-time simulations where the unitary or dissipative dynamics of interest is observed by tuning the Hamiltonian of the system~\cite{Altman2021}, its coupling with a reservoir and the properties of the reservoir~\cite{Harrington2022}. In this case, the exponential complexity of simulating a many-body system is built into the Hamiltonian itself and its stability in experiments imposes an upper limit on the observational time-range. The second comprises linear interferometer simulators~\cite{Knill2001,Carolan2015} where a single-photon unitary over $N$ modes is realised through $N(N-1)/2$ beam-splitters and phase-shifters~\cite{Reck1994Experimental,Clements2016Optimal}. Combined with unitary dilation and multi-photon input states, such programmable unitary~\cite{Carolan2015} can be used to simulate non-unitary~\cite{Nicola2022} or Lindblad dynamics~\cite{Sparrow2018-Nature} and quantum complexity problems~\cite{Tillmann2013,Spagnolo2014,Carolan2014}. In this section, we will show that a non-Hermitian Hamiltonian $H_\mathrm{eff}$ can be simulated by using nonlinear elements combined with reconfigurable-unitary~\cite{Carolan2015} circuit. 

The symplectic matrix $M=\exp(-iH_\mathrm{eff}t)$ on the operator space $\mathcal{S}$ of an $N$-mode system is parameterised by two complex, $N\times N$ matrices $F,G$ as
\begin{align}
\ket{A(t)}=M(t)\ket{A(0)}=
\left[\begin{array}{cc}
      F(t)   &  G(t) \\
      G^*(t)   &  F^*(t)
    \end{array}\right]    
    \left[\begin{array}{c}
    a^\dagger_j\\
    a_j
    \end{array}\right].
\end{align}
It translates into transformations on the Fock space as $a^\dagger_j(t)=U(t) a^\dagger_j U^\dagger(t)$ and $a(t)=U(t)aU^\dagger(t)$ where $U(t)=\exp(-iHt)$ is the unitary operator generated by the Hermitian, non-number-preserving Hamiltonian, Eq.\eqref{eq:H1}. To create a nonlinear optical circuit that simulates a non-Hermitian $H_\mathrm{eff}$ or equivalently, a non-unitary $M$, we use the singular value decomposition (SVD)~\cite{Horn1985,Tischler2018Arbitrary} that acts on the $\mathcal{S}$ space and the Fock-space as follows:
\begin{align}
\label{eq:symplectic}
    M(t)&=M_UM_\Sigma M_{V^\dagger},\\
    \label{eq:unitary}    
    a^\dagger_j(t)\ket{0}&=U\Sigma V^\dagger a^\dagger_j V\Sigma^\dagger U^\dagger\ket{0}.
\end{align}
Here, $M_U,M_{V^\dagger}$ are $2N\times 2N$ unitary matrices; their equivalent unitary operators $U,V^\dagger$ on the Fock-space are generated by {\it photon-number preserving} Hamiltonians, i.e. linear interferometers~\cite{Reck1994Experimental,Clements2016Optimal}; they leave the vacuum state unchanged, $U^\dagger\ket{0}=\ket{0}$. As linear optical circuits are based on beam-splitters and phase-shifters, and not on continuous-time Hamiltonian evolution, we drop the time-argument in equations from now on. The diagonal matrix $M_\Sigma$ has eigenvalues of the positive-definite matrix $MM^\dagger$ --- the singular values of $M$~\cite{Horn1985}. Its corresponding action on the Fock space, $a^\dagger\rightarrow \Sigma a^\dagger\Sigma^\dagger$ is generated by {\it photon-number changing} Hamiltonians. These operations, therefore, do not leave the vacuum unchanged, $\Sigma^\dagger\ket{0}\neq\ket{0}$. Finally, simulation of the non-unitary single-photon creation operator $a^\dagger_j$ requires a non-unitary process, such as two-mode squeezing followed by a heralding measurement. Specifically, let us consider the state generated by the two-mode squeezing Hamiltonian,  Eq.\eqref{eq:tms}, with modes $a_j,b$ and $\omega_j=0=\omega_b$,
\begin{align}
\ket{\psi}=e^{-iH_\mathrm{tms}t}\ket{0}\approx \ket{0}_j\ket{0}_b-(\kappa t)\ket{1}_j\ket{1}_b+\cdots
\end{align}
where we have ignored higher-order factors in $\kappa t$. This approximation is valid at low squeezing $\kappa t\ll 1$. When this state is heralded on mode $b$ with a single-photon detector, the resulting state $\ket{1}_j=a^\dagger_j\ket{0}$ is single-photon source for mode $j$.

Figure~\ref{fig:diagram} shows the schematic of a nonlinear optical circuit for implementing the Fock-space unitary \eqref{eq:unitary} corresponding to a symplectic transformation $M$ \eqref{eq:symplectic}. Reading from the left, the single-mode squeezer implements the diagonal unitary $\Sigma^\dagger$; a Clements scheme~\cite{Clements2016Optimal} implements the unitary operator $V$; a two-mode squeezer (red) heralded with the $b$-modes (green) implements the $a^\dagger$ operator; it is followed by $V^\dagger$, another single-mode squeezer $\Sigma$, and finally, the unitary operator $U$. Since the vacuum is invariant under the $U^\dagger$ operator, it is absent left of the first single-mode squeezer $\Sigma^\dagger$. With integrated, reconfigurable, universal linear optical circuits~\cite{Carolan2015,Sparrow2018-Nature} and photon sources, the SVD method provides a new avenue to realise non-Hermitian dynamics in the quantum domain. It can efficiently implemented using methods from Hamiltonian simulation algorithms, such as the Lie-Trotter-Suzuki decomposition~\cite{Barthel2020,Clinton2021,Childs2021} for Hamiltonians that can be expressed as sum of local Hamiltonians.


\section{Discussion}
\label{sec:disc}
Coherent, non-unitary dynamics generated by non-Hermitian Hamiltonians with antilinear symmetries~\cite{Ruzicka2021} ($\mathcal{PT}$ symmetry, anti-$\mathcal{PT}$ symmetry) has been extensively investigated in classical platforms where balanced gain and loss are easily implemented. However, their quantum realisations face fundamental challenges that include deleterious quantum~\cite{Caves1982} and thermal fluctuations~\cite{Kubo1966} and exponential-in-time cost of successful post-selection~\cite{Naghiloo2019,Quinn2023}. Alternatively, simulating the resultant non-unitary, trace-preserving quantum maps~\cite{Broadbridge1981} via Naimark or unitary dilation requires time-dependent Hermitian Hamiltonian~\cite{Gunther2008-proposal,halmos-1950,Nicola2022}. 

Here, we have shown that quadratic bosonic Hermitian Hamiltonians on the Fock space give rise to non-Hermitian Hamiltonians on the operator space. This mapping between canonical commutation relation preserving but boson-number-violating symplectic transformations~\cite{Balian1969,Colpa1978,Tsallis1978} and non-Hermitian Hamiltonians with antilinear symmetries provides a new way to simulate them using nonlinear optical transformations. We used this to propose an SVD scheme that can simulate arbitrary non-unitary transformations albeit with exponential resources as is true with currently used methods of simulating unitary evolution in linear optics~\cite{Knill2001,Carolan2015}. The idea of using squeezing, nonlinearity, or parametric driving to realise non-Hermitian dynamics has been proposed, over the years, in a number of classical~\cite{Antonosyan2015,Suchkov2016,Miri2016,Longhi2018,Humire2023} and quantum~\cite{Jiang2019,AshClerk2019,Luo2022,Gaikwad2023} platforms. Our analysis provides a unified treatment that is applicable to the quantum photonic and phononic~\cite{Burd2021} domains.  


\bibliography{PTSqueezing,ptyj1,ptyj2}

\begin{thebibliography}{69}%
\makeatletter
\providecommand \@ifxundefined [1]{%
 \@ifx{#1\undefined}
}%
\providecommand \@ifnum [1]{%
 \ifnum #1\expandafter \@firstoftwo
 \else \expandafter \@secondoftwo
 \fi
}%
\providecommand \@ifx [1]{%
 \ifx #1\expandafter \@firstoftwo
 \else \expandafter \@secondoftwo
 \fi
}%
\providecommand \natexlab [1]{#1}%
\providecommand \enquote  [1]{``#1''}%
\providecommand \bibnamefont  [1]{#1}%
\providecommand \bibfnamefont [1]{#1}%
\providecommand \citenamefont [1]{#1}%
\providecommand \href@noop [0]{\@secondoftwo}%
\providecommand \href [0]{\begingroup \@sanitize@url \@href}%
\providecommand \@href[1]{\@@startlink{#1}\@@href}%
\providecommand \@@href[1]{\endgroup#1\@@endlink}%
\providecommand \@sanitize@url [0]{\catcode `\\12\catcode `\$12\catcode
  `\&12\catcode `\#12\catcode `\^12\catcode `\_12\catcode `\%12\relax}%
\providecommand \@@startlink[1]{}%
\providecommand \@@endlink[0]{}%
\providecommand \url  [0]{\begingroup\@sanitize@url \@url }%
\providecommand \@url [1]{\endgroup\@href {#1}{\urlprefix }}%
\providecommand \urlprefix  [0]{URL }%
\providecommand \Eprint [0]{\href }%
\providecommand \doibase [0]{http://dx.doi.org/}%
\providecommand \selectlanguage [0]{\@gobble}%
\providecommand \bibinfo  [0]{\@secondoftwo}%
\providecommand \bibfield  [0]{\@secondoftwo}%
\providecommand \translation [1]{[#1]}%
\providecommand \BibitemOpen [0]{}%
\providecommand \bibitemStop [0]{}%
\providecommand \bibitemNoStop [0]{.\EOS\space}%
\providecommand \EOS [0]{\spacefactor3000\relax}%
\providecommand \BibitemShut  [1]{\csname bibitem#1\endcsname}%
\let\auto@bib@innerbib\@empty
\bibitem [{\citenamefont {Bender}\ and\ \citenamefont
  {Boettcher}(1998)}]{Bender1998}%
  \BibitemOpen
  \bibfield  {author} {\bibinfo {author} {\bibfnamefont {C.~M.}\ \bibnamefont
  {Bender}}\ and\ \bibinfo {author} {\bibfnamefont {S.}~\bibnamefont
  {Boettcher}},\ }\bibfield  {title} {\enquote {\bibinfo {title} {Real spectra
  in non-hermitian hamiltonians having $\mathcal{PT}$ symmetry},}\ }\href
  {\doibase 10.1103/PhysRevLett.80.5243} {\bibfield  {journal} {\bibinfo
  {journal} {Phys. Rev. Lett.}\ }\textbf {\bibinfo {volume} {80}},\ \bibinfo
  {pages} {5243--5246} (\bibinfo {year} {1998})}\BibitemShut {NoStop}%
\bibitem [{\citenamefont {Bender}, \citenamefont {Brody},\ and\ \citenamefont
  {Jones}(2002)}]{Bender2001}%
  \BibitemOpen
  \bibfield  {author} {\bibinfo {author} {\bibfnamefont {C.~M.}\ \bibnamefont
  {Bender}}, \bibinfo {author} {\bibfnamefont {D.~C.}\ \bibnamefont {Brody}}, \
  and\ \bibinfo {author} {\bibfnamefont {H.~F.}\ \bibnamefont {Jones}},\
  }\bibfield  {title} {\enquote {\bibinfo {title} {Complex extension of quantum
  mechanics},}\ }\href {\doibase 10.1103/PhysRevLett.89.270401} {\bibfield
  {journal} {\bibinfo  {journal} {Phys. Rev. Lett.}\ }\textbf {\bibinfo
  {volume} {89}},\ \bibinfo {pages} {270401} (\bibinfo {year}
  {2002})}\BibitemShut {NoStop}%
\bibitem [{\citenamefont {Mostafazadeh}(2002)}]{Mostafazadeh2002}%
  \BibitemOpen
  \bibfield  {author} {\bibinfo {author} {\bibfnamefont {A.}~\bibnamefont
  {Mostafazadeh}},\ }\bibfield  {title} {\enquote {\bibinfo {title}
  {Pseudo-hermiticity versus {PT} symmetry: The necessary condition for the
  reality of the spectrum of a non-hermitian hamiltonian},}\ }\href {\doibase
  10.1063/1.1418246} {\bibfield  {journal} {\bibinfo  {journal} {Journal of
  Mathematical Physics}\ }\textbf {\bibinfo {volume} {43}},\ \bibinfo {pages}
  {205--214} (\bibinfo {year} {2002})}\BibitemShut {NoStop}%
\bibitem [{\citenamefont {Mostafazadeh}(2003)}]{Mostafazadeh2003-equivalence}%
  \BibitemOpen
  \bibfield  {author} {\bibinfo {author} {\bibfnamefont {A.}~\bibnamefont
  {Mostafazadeh}},\ }\bibfield  {title} {\enquote {\bibinfo {title} {{Exact
  PT-symmetry is equivalent to Hermiticity}},}\ }\href {\doibase
  10.1088/0305-4470/36/25/312} {\bibfield  {journal} {\bibinfo  {journal}
  {Journal of Physics A: Mathematical and General}\ }\textbf {\bibinfo {volume}
  {36}},\ \bibinfo {pages} {7081--7091} (\bibinfo {year} {2003})}\BibitemShut
  {NoStop}%
\bibitem [{\citenamefont {Feng}, \citenamefont {El-Ganainy},\ and\
  \citenamefont {Ge}(2017)}]{Feng2017}%
  \BibitemOpen
  \bibfield  {author} {\bibinfo {author} {\bibfnamefont {L.}~\bibnamefont
  {Feng}}, \bibinfo {author} {\bibfnamefont {R.}~\bibnamefont {El-Ganainy}}, \
  and\ \bibinfo {author} {\bibfnamefont {L.}~\bibnamefont {Ge}},\ }\bibfield
  {title} {\enquote {\bibinfo {title} {Non-hermitian photonics based on
  parity{\textendash}time symmetry},}\ }\href {\doibase
  10.1038/s41566-017-0031-1} {\bibfield  {journal} {\bibinfo  {journal} {Nature
  Photonics}\ }\textbf {\bibinfo {volume} {11}},\ \bibinfo {pages} {752--762}
  (\bibinfo {year} {2017})}\BibitemShut {NoStop}%
\bibitem [{\citenamefont {{\"O}zdemir}\ \emph {et~al.}(2019)\citenamefont
  {{\"O}zdemir}, \citenamefont {Rotter}, \citenamefont {Nori},\ and\
  \citenamefont {Yang}}]{Ozdemir2019}%
  \BibitemOpen
  \bibfield  {author} {\bibinfo {author} {\bibfnamefont {{\c{S}}.~K.}\
  \bibnamefont {{\"O}zdemir}}, \bibinfo {author} {\bibfnamefont
  {S.}~\bibnamefont {Rotter}}, \bibinfo {author} {\bibfnamefont
  {F.}~\bibnamefont {Nori}}, \ and\ \bibinfo {author} {\bibfnamefont
  {L.}~\bibnamefont {Yang}},\ }\bibfield  {title} {\enquote {\bibinfo {title}
  {Parity--time symmetry and exceptional points in photonics},}\ }\href
  {\doibase 10.1038/s41563-019-0304-9} {\bibfield  {journal} {\bibinfo
  {journal} {Nature Materials}\ }\textbf {\bibinfo {volume} {18}},\ \bibinfo
  {pages} {783--798} (\bibinfo {year} {2019})}\BibitemShut {NoStop}%
\bibitem [{\citenamefont {Kato}(1995)}]{Kato1995}%
  \BibitemOpen
  \bibfield  {author} {\bibinfo {author} {\bibfnamefont {T.}~\bibnamefont
  {Kato}},\ }\href {\doibase 10.1007/978-3-642-66282-9} {\emph {\bibinfo
  {title} {Perturbation Theory for Linear Operators}}}\ (\bibinfo  {publisher}
  {Springer Berlin Heidelberg},\ \bibinfo {year} {1995})\BibitemShut {NoStop}%
\bibitem [{\citenamefont {Miri}\ and\ \citenamefont
  {Al{\`{u}}}(2019)}]{Miri2019}%
  \BibitemOpen
  \bibfield  {author} {\bibinfo {author} {\bibfnamefont {M.-A.}\ \bibnamefont
  {Miri}}\ and\ \bibinfo {author} {\bibfnamefont {A.}~\bibnamefont
  {Al{\`{u}}}},\ }\bibfield  {title} {\enquote {\bibinfo {title} {Exceptional
  points in optics and photonics},}\ }\href {\doibase 10.1126/science.aar7709}
  {\bibfield  {journal} {\bibinfo  {journal} {Science}\ }\textbf {\bibinfo
  {volume} {363}},\ \bibinfo {pages} {eaar7709} (\bibinfo {year}
  {2019})}\BibitemShut {NoStop}%
\bibitem [{\citenamefont {Regensburger}\ \emph {et~al.}(2012)\citenamefont
  {Regensburger}, \citenamefont {Bersch}, \citenamefont {Miri}, \citenamefont
  {Onishchukov}, \citenamefont {Christodoulides},\ and\ \citenamefont
  {Peschel}}]{Regensburger2012}%
  \BibitemOpen
  \bibfield  {author} {\bibinfo {author} {\bibfnamefont {A.}~\bibnamefont
  {Regensburger}}, \bibinfo {author} {\bibfnamefont {C.}~\bibnamefont
  {Bersch}}, \bibinfo {author} {\bibfnamefont {M.-A.}\ \bibnamefont {Miri}},
  \bibinfo {author} {\bibfnamefont {G.}~\bibnamefont {Onishchukov}}, \bibinfo
  {author} {\bibfnamefont {D.~N.}\ \bibnamefont {Christodoulides}}, \ and\
  \bibinfo {author} {\bibfnamefont {U.}~\bibnamefont {Peschel}},\ }\bibfield
  {title} {\enquote {\bibinfo {title} {Parity{\textendash}time synthetic
  photonic lattices},}\ }\href {\doibase 10.1038/nature11298} {\bibfield
  {journal} {\bibinfo  {journal} {Nature}\ }\textbf {\bibinfo {volume} {488}},\
  \bibinfo {pages} {167--171} (\bibinfo {year} {2012})}\BibitemShut {NoStop}%
\bibitem [{\citenamefont {Hodaei}\ \emph {et~al.}(2014)\citenamefont {Hodaei},
  \citenamefont {Miri}, \citenamefont {Heinrich}, \citenamefont
  {Christodoulides},\ and\ \citenamefont {Khajavikhan}}]{Hodaei2014}%
  \BibitemOpen
  \bibfield  {author} {\bibinfo {author} {\bibfnamefont {H.}~\bibnamefont
  {Hodaei}}, \bibinfo {author} {\bibfnamefont {M.-A.}\ \bibnamefont {Miri}},
  \bibinfo {author} {\bibfnamefont {M.}~\bibnamefont {Heinrich}}, \bibinfo
  {author} {\bibfnamefont {D.~N.}\ \bibnamefont {Christodoulides}}, \ and\
  \bibinfo {author} {\bibfnamefont {M.}~\bibnamefont {Khajavikhan}},\
  }\bibfield  {title} {\enquote {\bibinfo {title} {Parity-time-symmetric
  microring lasers},}\ }\href {\doibase 10.1126/science.1258480} {\bibfield
  {journal} {\bibinfo  {journal} {Science}\ }\textbf {\bibinfo {volume}
  {346}},\ \bibinfo {pages} {975--978} (\bibinfo {year} {2014})}\BibitemShut
  {NoStop}%
\bibitem [{\citenamefont {Peng}\ \emph
  {et~al.}(2014{\natexlab{a}})\citenamefont {Peng}, \citenamefont
  {\"{O}zdemir}, \citenamefont {Lei}, \citenamefont {Monifi}, \citenamefont
  {Gianfreda}, \citenamefont {Long}, \citenamefont {Fan}, \citenamefont {Nori},
  \citenamefont {Bender},\ and\ \citenamefont {Yang}}]{Peng2014}%
  \BibitemOpen
  \bibfield  {author} {\bibinfo {author} {\bibfnamefont {B.}~\bibnamefont
  {Peng}}, \bibinfo {author} {\bibfnamefont {{\c{S}}.~K.}\ \bibnamefont
  {\"{O}zdemir}}, \bibinfo {author} {\bibfnamefont {F.}~\bibnamefont {Lei}},
  \bibinfo {author} {\bibfnamefont {F.}~\bibnamefont {Monifi}}, \bibinfo
  {author} {\bibfnamefont {M.}~\bibnamefont {Gianfreda}}, \bibinfo {author}
  {\bibfnamefont {G.~L.}\ \bibnamefont {Long}}, \bibinfo {author}
  {\bibfnamefont {S.}~\bibnamefont {Fan}}, \bibinfo {author} {\bibfnamefont
  {F.}~\bibnamefont {Nori}}, \bibinfo {author} {\bibfnamefont {C.~M.}\
  \bibnamefont {Bender}}, \ and\ \bibinfo {author} {\bibfnamefont
  {L.}~\bibnamefont {Yang}},\ }\bibfield  {title} {\enquote {\bibinfo {title}
  {Parity{\textendash}time-symmetric whispering-gallery microcavities},}\
  }\href {\doibase 10.1038/nphys2927} {\bibfield  {journal} {\bibinfo
  {journal} {Nature Physics}\ }\textbf {\bibinfo {volume} {10}},\ \bibinfo
  {pages} {394--398} (\bibinfo {year} {2014}{\natexlab{a}})}\BibitemShut
  {NoStop}%
\bibitem [{\citenamefont {Peng}\ \emph
  {et~al.}(2014{\natexlab{b}})\citenamefont {Peng}, \citenamefont
  {\"{O}zdemir}, \citenamefont {Rotter}, \citenamefont {Yilmaz}, \citenamefont
  {Liertzer}, \citenamefont {Monifi}, \citenamefont {Bender}, \citenamefont
  {Nori},\ and\ \citenamefont {Yang}}]{Peng2014b}%
  \BibitemOpen
  \bibfield  {author} {\bibinfo {author} {\bibfnamefont {B.}~\bibnamefont
  {Peng}}, \bibinfo {author} {\bibfnamefont {{\c{S}}.~K.}\ \bibnamefont
  {\"{O}zdemir}}, \bibinfo {author} {\bibfnamefont {S.}~\bibnamefont {Rotter}},
  \bibinfo {author} {\bibfnamefont {H.}~\bibnamefont {Yilmaz}}, \bibinfo
  {author} {\bibfnamefont {M.}~\bibnamefont {Liertzer}}, \bibinfo {author}
  {\bibfnamefont {F.}~\bibnamefont {Monifi}}, \bibinfo {author} {\bibfnamefont
  {C.~M.}\ \bibnamefont {Bender}}, \bibinfo {author} {\bibfnamefont
  {F.}~\bibnamefont {Nori}}, \ and\ \bibinfo {author} {\bibfnamefont
  {L.}~\bibnamefont {Yang}},\ }\bibfield  {title} {\enquote {\bibinfo {title}
  {Loss-induced suppression and revival of lasing},}\ }\href {\doibase
  10.1126/science.1258004} {\bibfield  {journal} {\bibinfo  {journal}
  {Science}\ }\textbf {\bibinfo {volume} {346}},\ \bibinfo {pages} {328--332}
  (\bibinfo {year} {2014}{\natexlab{b}})}\BibitemShut {NoStop}%
\bibitem [{\citenamefont {Feng}\ \emph {et~al.}(2014)\citenamefont {Feng},
  \citenamefont {Wong}, \citenamefont {Ma}, \citenamefont {Wang},\ and\
  \citenamefont {Zhang}}]{Feng2014}%
  \BibitemOpen
  \bibfield  {author} {\bibinfo {author} {\bibfnamefont {L.}~\bibnamefont
  {Feng}}, \bibinfo {author} {\bibfnamefont {Z.~J.}\ \bibnamefont {Wong}},
  \bibinfo {author} {\bibfnamefont {R.-M.}\ \bibnamefont {Ma}}, \bibinfo
  {author} {\bibfnamefont {Y.}~\bibnamefont {Wang}}, \ and\ \bibinfo {author}
  {\bibfnamefont {X.}~\bibnamefont {Zhang}},\ }\bibfield  {title} {\enquote
  {\bibinfo {title} {Single-mode laser by parity-time symmetry breaking},}\
  }\href {\doibase 10.1126/science.1258479} {\bibfield  {journal} {\bibinfo
  {journal} {Science}\ }\textbf {\bibinfo {volume} {346}},\ \bibinfo {pages}
  {972--975} (\bibinfo {year} {2014})}\BibitemShut {NoStop}%
\bibitem [{\citenamefont {Zhu}\ \emph {et~al.}(2014)\citenamefont {Zhu},
  \citenamefont {Ramezani}, \citenamefont {Shi}, \citenamefont {Zhu},\ and\
  \citenamefont {Zhang}}]{Zhu2014}%
  \BibitemOpen
  \bibfield  {author} {\bibinfo {author} {\bibfnamefont {X.}~\bibnamefont
  {Zhu}}, \bibinfo {author} {\bibfnamefont {H.}~\bibnamefont {Ramezani}},
  \bibinfo {author} {\bibfnamefont {C.}~\bibnamefont {Shi}}, \bibinfo {author}
  {\bibfnamefont {J.}~\bibnamefont {Zhu}}, \ and\ \bibinfo {author}
  {\bibfnamefont {X.}~\bibnamefont {Zhang}},\ }\bibfield  {title} {\enquote
  {\bibinfo {title} {$\mathcal{P}\mathcal{T}$-symmetric acoustics},}\ }\href
  {\doibase 10.1103/PhysRevX.4.031042} {\bibfield  {journal} {\bibinfo
  {journal} {Phys. Rev. X}\ }\textbf {\bibinfo {volume} {4}},\ \bibinfo {pages}
  {031042} (\bibinfo {year} {2014})}\BibitemShut {NoStop}%
\bibitem [{\citenamefont {Schindler}\ \emph {et~al.}(2011)\citenamefont
  {Schindler}, \citenamefont {Li}, \citenamefont {Zheng}, \citenamefont
  {Ellis},\ and\ \citenamefont {Kottos}}]{Schindler2011}%
  \BibitemOpen
  \bibfield  {author} {\bibinfo {author} {\bibfnamefont {J.}~\bibnamefont
  {Schindler}}, \bibinfo {author} {\bibfnamefont {A.}~\bibnamefont {Li}},
  \bibinfo {author} {\bibfnamefont {M.~C.}\ \bibnamefont {Zheng}}, \bibinfo
  {author} {\bibfnamefont {F.~M.}\ \bibnamefont {Ellis}}, \ and\ \bibinfo
  {author} {\bibfnamefont {T.}~\bibnamefont {Kottos}},\ }\bibfield  {title}
  {\enquote {\bibinfo {title} {{Experimental study of active LRC circuits with
  PT symmetries}},}\ }\href {\doibase 10.1103/PhysRevA.84.040101} {\bibfield
  {journal} {\bibinfo  {journal} {Physical Review A - Atomic, Molecular, and
  Optical Physics}\ }\textbf {\bibinfo {volume} {84}},\ \bibinfo {pages} {1--5}
  (\bibinfo {year} {2011})}\BibitemShut {NoStop}%
\bibitem [{\citenamefont {Wang}\ \emph {et~al.}(2020)\citenamefont {Wang},
  \citenamefont {Fang}, \citenamefont {Xie}, \citenamefont {Dong},
  \citenamefont {Joglekar}, \citenamefont {Wang}, \citenamefont {Li},\ and\
  \citenamefont {Luo}}]{Wang2020}%
  \BibitemOpen
  \bibfield  {author} {\bibinfo {author} {\bibfnamefont {T.}~\bibnamefont
  {Wang}}, \bibinfo {author} {\bibfnamefont {J.}~\bibnamefont {Fang}}, \bibinfo
  {author} {\bibfnamefont {Z.}~\bibnamefont {Xie}}, \bibinfo {author}
  {\bibfnamefont {N.}~\bibnamefont {Dong}}, \bibinfo {author} {\bibfnamefont
  {Y.~N.}\ \bibnamefont {Joglekar}}, \bibinfo {author} {\bibfnamefont
  {Z.}~\bibnamefont {Wang}}, \bibinfo {author} {\bibfnamefont {J.}~\bibnamefont
  {Li}}, \ and\ \bibinfo {author} {\bibfnamefont {L.}~\bibnamefont {Luo}},\
  }\bibfield  {title} {\enquote {\bibinfo {title} {Observation of two pt
  transitions in an electric circuit with balanced gain and loss},}\ }\href
  {\doibase 10.1140/epjd/e2020-10131-7} {\bibfield  {journal} {\bibinfo
  {journal} {The European Physical Journal D}\ }\textbf {\bibinfo {volume}
  {74}} (\bibinfo {year} {2020}),\ 10.1140/epjd/e2020-10131-7}\BibitemShut
  {NoStop}%
\bibitem [{\citenamefont {Quiroz-Ju\'arez}\ \emph {et~al.}(2022)\citenamefont
  {Quiroz-Ju\'arez}, \citenamefont {Agarwal}, \citenamefont {Cochran},
  \citenamefont {Arag\'on}, \citenamefont {Joglekar},\ and\ \citenamefont
  {Le\'on-Montiel}}]{Quiroz2022}%
  \BibitemOpen
  \bibfield  {author} {\bibinfo {author} {\bibfnamefont {M.~A.}\ \bibnamefont
  {Quiroz-Ju\'arez}}, \bibinfo {author} {\bibfnamefont {K.~S.}\ \bibnamefont
  {Agarwal}}, \bibinfo {author} {\bibfnamefont {Z.~A.}\ \bibnamefont
  {Cochran}}, \bibinfo {author} {\bibfnamefont {J.~L.}\ \bibnamefont
  {Arag\'on}}, \bibinfo {author} {\bibfnamefont {Y.~N.}\ \bibnamefont
  {Joglekar}}, \ and\ \bibinfo {author} {\bibfnamefont {R.~d.~J.}\ \bibnamefont
  {Le\'on-Montiel}},\ }\bibfield  {title} {\enquote {\bibinfo {title}
  {On-demand parity-time symmetry in a lone oscillator through complex
  synthetic gauge fields},}\ }\href {\doibase 10.1103/PhysRevApplied.18.054034}
  {\bibfield  {journal} {\bibinfo  {journal} {Phys. Rev. Appl.}\ }\textbf
  {\bibinfo {volume} {18}},\ \bibinfo {pages} {054034} (\bibinfo {year}
  {2022})}\BibitemShut {NoStop}%
\bibitem [{\citenamefont {Li}\ \emph {et~al.}(2019)\citenamefont {Li},
  \citenamefont {Harter}, \citenamefont {Liu}, \citenamefont {de~Melo},
  \citenamefont {Joglekar},\ and\ \citenamefont {Luo}}]{Li2019}%
  \BibitemOpen
  \bibfield  {author} {\bibinfo {author} {\bibfnamefont {J.}~\bibnamefont
  {Li}}, \bibinfo {author} {\bibfnamefont {A.~K.}\ \bibnamefont {Harter}},
  \bibinfo {author} {\bibfnamefont {J.}~\bibnamefont {Liu}}, \bibinfo {author}
  {\bibfnamefont {L.}~\bibnamefont {de~Melo}}, \bibinfo {author} {\bibfnamefont
  {Y.~N.}\ \bibnamefont {Joglekar}}, \ and\ \bibinfo {author} {\bibfnamefont
  {L.}~\bibnamefont {Luo}},\ }\bibfield  {title} {\enquote {\bibinfo {title}
  {Observation of parity-time symmetry breaking transitions in a dissipative
  floquet system of ultracold atoms},}\ }\href {\doibase
  10.1038/s41467-019-08596-1} {\bibfield  {journal} {\bibinfo  {journal}
  {Nature Communications}\ }\textbf {\bibinfo {volume} {10}} (\bibinfo {year}
  {2019}),\ 10.1038/s41467-019-08596-1}\BibitemShut {NoStop}%
\bibitem [{\citenamefont {Wu}\ \emph {et~al.}(2019)\citenamefont {Wu},
  \citenamefont {Liu}, \citenamefont {Geng}, \citenamefont {Song},
  \citenamefont {Ye}, \citenamefont {Duan}, \citenamefont {Rong},\ and\
  \citenamefont {Du}}]{Wu2019}%
  \BibitemOpen
  \bibfield  {author} {\bibinfo {author} {\bibfnamefont {Y.}~\bibnamefont
  {Wu}}, \bibinfo {author} {\bibfnamefont {W.}~\bibnamefont {Liu}}, \bibinfo
  {author} {\bibfnamefont {J.}~\bibnamefont {Geng}}, \bibinfo {author}
  {\bibfnamefont {X.}~\bibnamefont {Song}}, \bibinfo {author} {\bibfnamefont
  {X.}~\bibnamefont {Ye}}, \bibinfo {author} {\bibfnamefont {C.-K.}\
  \bibnamefont {Duan}}, \bibinfo {author} {\bibfnamefont {X.}~\bibnamefont
  {Rong}}, \ and\ \bibinfo {author} {\bibfnamefont {J.}~\bibnamefont {Du}},\
  }\bibfield  {title} {\enquote {\bibinfo {title} {Observation of parity-time
  symmetry breaking in a single-spin system},}\ }\href {\doibase
  10.1126/science.aaw8205} {\bibfield  {journal} {\bibinfo  {journal}
  {Science}\ }\textbf {\bibinfo {volume} {364}},\ \bibinfo {pages} {878--880}
  (\bibinfo {year} {2019})}\BibitemShut {NoStop}%
\bibitem [{\citenamefont {Klauck}\ \emph {et~al.}(2019)\citenamefont {Klauck},
  \citenamefont {Teuber}, \citenamefont {Ornigotti}, \citenamefont {Heinrich},
  \citenamefont {Scheel},\ and\ \citenamefont {Szameit}}]{Klauck2019}%
  \BibitemOpen
  \bibfield  {author} {\bibinfo {author} {\bibfnamefont {F.}~\bibnamefont
  {Klauck}}, \bibinfo {author} {\bibfnamefont {L.}~\bibnamefont {Teuber}},
  \bibinfo {author} {\bibfnamefont {M.}~\bibnamefont {Ornigotti}}, \bibinfo
  {author} {\bibfnamefont {M.}~\bibnamefont {Heinrich}}, \bibinfo {author}
  {\bibfnamefont {S.}~\bibnamefont {Scheel}}, \ and\ \bibinfo {author}
  {\bibfnamefont {A.}~\bibnamefont {Szameit}},\ }\bibfield  {title} {\enquote
  {\bibinfo {title} {Observation of $\mathcal{PT}$-symmetric quantum
  interference},}\ }\href {\doibase 10.1038/s41566-019-0517-0} {\bibfield
  {journal} {\bibinfo  {journal} {Nature Photonics}\ }\textbf {\bibinfo
  {volume} {13}},\ \bibinfo {pages} {883--887} (\bibinfo {year}
  {2019})}\BibitemShut {NoStop}%
\bibitem [{\citenamefont {Naghiloo}\ \emph {et~al.}(2019)\citenamefont
  {Naghiloo}, \citenamefont {Abbasi}, \citenamefont {Joglekar},\ and\
  \citenamefont {Murch}}]{Naghiloo2019}%
  \BibitemOpen
  \bibfield  {author} {\bibinfo {author} {\bibfnamefont {M.}~\bibnamefont
  {Naghiloo}}, \bibinfo {author} {\bibfnamefont {M.}~\bibnamefont {Abbasi}},
  \bibinfo {author} {\bibfnamefont {Y.~N.}\ \bibnamefont {Joglekar}}, \ and\
  \bibinfo {author} {\bibfnamefont {K.~W.}\ \bibnamefont {Murch}},\ }\bibfield
  {title} {\enquote {\bibinfo {title} {Quantum state tomography across the
  exceptional point in a single dissipative qubit},}\ }\href {\doibase
  10.1038/s41567-019-0652-z} {\bibfield  {journal} {\bibinfo  {journal} {Nature
  Physics}\ }\textbf {\bibinfo {volume} {15}},\ \bibinfo {pages} {1232--1236}
  (\bibinfo {year} {2019})}\BibitemShut {NoStop}%
\bibitem [{\citenamefont {Ding}\ \emph {et~al.}(2021)\citenamefont {Ding},
  \citenamefont {Shi}, \citenamefont {Zhang}, \citenamefont {Shen},
  \citenamefont {Zhang},\ and\ \citenamefont {Zhang}}]{Ding2021}%
  \BibitemOpen
  \bibfield  {author} {\bibinfo {author} {\bibfnamefont {L.}~\bibnamefont
  {Ding}}, \bibinfo {author} {\bibfnamefont {K.}~\bibnamefont {Shi}}, \bibinfo
  {author} {\bibfnamefont {Q.}~\bibnamefont {Zhang}}, \bibinfo {author}
  {\bibfnamefont {D.}~\bibnamefont {Shen}}, \bibinfo {author} {\bibfnamefont
  {X.}~\bibnamefont {Zhang}}, \ and\ \bibinfo {author} {\bibfnamefont
  {W.}~\bibnamefont {Zhang}},\ }\bibfield  {title} {\enquote {\bibinfo {title}
  {Experimental determination of $\mathcal{P}\mathcal{T}$-symmetric exceptional
  points in a single trapped ion},}\ }\href {\doibase
  10.1103/PhysRevLett.126.083604} {\bibfield  {journal} {\bibinfo  {journal}
  {Phys. Rev. Lett.}\ }\textbf {\bibinfo {volume} {126}},\ \bibinfo {pages}
  {083604} (\bibinfo {year} {2021})}\BibitemShut {NoStop}%
\bibitem [{\citenamefont {Kubo}(1966)}]{Kubo1966}%
  \BibitemOpen
  \bibfield  {author} {\bibinfo {author} {\bibfnamefont {R.}~\bibnamefont
  {Kubo}},\ }\bibfield  {title} {\enquote {\bibinfo {title} {The
  fluctuation-dissipation theorem},}\ }\href {\doibase
  10.1088/0034-4885/29/1/306} {\bibfield  {journal} {\bibinfo  {journal}
  {Reports on Progress in Physics}\ }\textbf {\bibinfo {volume} {29}},\
  \bibinfo {pages} {255--284} (\bibinfo {year} {1966})}\BibitemShut {NoStop}%
\bibitem [{\citenamefont {Caves}(1982)}]{Caves1982}%
  \BibitemOpen
  \bibfield  {author} {\bibinfo {author} {\bibfnamefont {C.~M.}\ \bibnamefont
  {Caves}},\ }\bibfield  {title} {\enquote {\bibinfo {title} {Quantum limits on
  noise in linear amplifiers},}\ }\href {\doibase 10.1103/physrevd.26.1817}
  {\bibfield  {journal} {\bibinfo  {journal} {Physical Review D}\ }\textbf
  {\bibinfo {volume} {26}},\ \bibinfo {pages} {1817--1839} (\bibinfo {year}
  {1982})}\BibitemShut {NoStop}%
\bibitem [{\citenamefont {Scheel}\ and\ \citenamefont
  {Szameit}(2018)}]{Scheel2018}%
  \BibitemOpen
  \bibfield  {author} {\bibinfo {author} {\bibfnamefont {S.}~\bibnamefont
  {Scheel}}\ and\ \bibinfo {author} {\bibfnamefont {A.}~\bibnamefont
  {Szameit}},\ }\bibfield  {title} {\enquote {\bibinfo {title}
  {$\mathcal{PT}$-symmetric photonic quantum systems with gain and loss do not
  exist},}\ }\href {\doibase 10.1209/0295-5075/122/34001} {\bibfield  {journal}
  {\bibinfo  {journal} {{EPL} (Europhysics Letters)}\ }\textbf {\bibinfo
  {volume} {122}},\ \bibinfo {pages} {34001} (\bibinfo {year}
  {2018})}\BibitemShut {NoStop}%
\bibitem [{\citenamefont {Guo}\ \emph {et~al.}(2009)\citenamefont {Guo},
  \citenamefont {Salamo}, \citenamefont {Duchesne}, \citenamefont {Morandotti},
  \citenamefont {Volatier-Ravat}, \citenamefont {Aimez}, \citenamefont
  {Siviloglou},\ and\ \citenamefont {Christodoulides}}]{guo09}%
  \BibitemOpen
  \bibfield  {author} {\bibinfo {author} {\bibfnamefont {A.}~\bibnamefont
  {Guo}}, \bibinfo {author} {\bibfnamefont {G.~J.}\ \bibnamefont {Salamo}},
  \bibinfo {author} {\bibfnamefont {D.}~\bibnamefont {Duchesne}}, \bibinfo
  {author} {\bibfnamefont {R.}~\bibnamefont {Morandotti}}, \bibinfo {author}
  {\bibfnamefont {M.}~\bibnamefont {Volatier-Ravat}}, \bibinfo {author}
  {\bibfnamefont {V.}~\bibnamefont {Aimez}}, \bibinfo {author} {\bibfnamefont
  {G.~A.}\ \bibnamefont {Siviloglou}}, \ and\ \bibinfo {author} {\bibfnamefont
  {D.~N.}\ \bibnamefont {Christodoulides}},\ }\bibfield  {title} {\enquote
  {\bibinfo {title} {Observation of $\mathcal{P}\mathcal{T}$-symmetry breaking
  in complex optical potentials},}\ }\href {\doibase
  10.1103/PhysRevLett.103.093902} {\bibfield  {journal} {\bibinfo  {journal}
  {Phys. Rev. Lett.}\ }\textbf {\bibinfo {volume} {103}},\ \bibinfo {pages}
  {093902} (\bibinfo {year} {2009})}\BibitemShut {NoStop}%
\bibitem [{\citenamefont {G{\"{u}}nther}\ and\ \citenamefont
  {Samsonov}(2008)}]{Gunther2008-proposal}%
  \BibitemOpen
  \bibfield  {author} {\bibinfo {author} {\bibfnamefont {U.}~\bibnamefont
  {G{\"{u}}nther}}\ and\ \bibinfo {author} {\bibfnamefont {B.~F.}\ \bibnamefont
  {Samsonov}},\ }\bibfield  {title} {\enquote {\bibinfo {title}
  {{Naimark-Dilated PT-Symmetric Brachistochrone}},}\ }\href {\doibase
  10.1103/PhysRevLett.101.230404} {\bibfield  {journal} {\bibinfo  {journal}
  {Physical Review Letters}\ }\textbf {\bibinfo {volume} {101}},\ \bibinfo
  {pages} {1--4} (\bibinfo {year} {2008})}\BibitemShut {NoStop}%
\bibitem [{\citenamefont {R.~Halmos}(1950)}]{halmos-1950}%
  \BibitemOpen
  \bibfield  {author} {\bibinfo {author} {\bibfnamefont {P.}~\bibnamefont
  {R.~Halmos}},\ }\bibfield  {title} {\enquote {\bibinfo {title} {Normal
  dilations and extensions of operators},}\ }\href {\doibase
  10.1007/978-1-4613-8208-9\_9} {\bibfield  {journal} {\bibinfo  {journal}
  {Summa Bras. Math.}\ }\textbf {\bibinfo {volume} {2}} (\bibinfo {year}
  {1950}),\ 10.1007/978-1-4613-8208-9\_9}\BibitemShut {NoStop}%
\bibitem [{\citenamefont {Maraviglia}\ \emph {et~al.}(2022)\citenamefont
  {Maraviglia}, \citenamefont {Yard}, \citenamefont {Wakefield}, \citenamefont
  {Carolan}, \citenamefont {Sparrow}, \citenamefont {Chakhmakhchyan},
  \citenamefont {Harrold}, \citenamefont {Hashimoto}, \citenamefont {Matsuda},
  \citenamefont {Harter}, \citenamefont {Joglekar},\ and\ \citenamefont
  {Laing}}]{Nicola2022}%
  \BibitemOpen
  \bibfield  {author} {\bibinfo {author} {\bibfnamefont {N.}~\bibnamefont
  {Maraviglia}}, \bibinfo {author} {\bibfnamefont {P.}~\bibnamefont {Yard}},
  \bibinfo {author} {\bibfnamefont {R.}~\bibnamefont {Wakefield}}, \bibinfo
  {author} {\bibfnamefont {J.}~\bibnamefont {Carolan}}, \bibinfo {author}
  {\bibfnamefont {C.}~\bibnamefont {Sparrow}}, \bibinfo {author} {\bibfnamefont
  {L.}~\bibnamefont {Chakhmakhchyan}}, \bibinfo {author} {\bibfnamefont
  {C.}~\bibnamefont {Harrold}}, \bibinfo {author} {\bibfnamefont
  {T.}~\bibnamefont {Hashimoto}}, \bibinfo {author} {\bibfnamefont
  {N.}~\bibnamefont {Matsuda}}, \bibinfo {author} {\bibfnamefont {A.~K.}\
  \bibnamefont {Harter}}, \bibinfo {author} {\bibfnamefont {Y.~N.}\
  \bibnamefont {Joglekar}}, \ and\ \bibinfo {author} {\bibfnamefont
  {A.}~\bibnamefont {Laing}},\ }\bibfield  {title} {\enquote {\bibinfo {title}
  {Photonic quantum simulations of coupled $\mathcal{PT}$-symmetric
  hamiltonians},}\ }\href {\doibase 10.1103/PhysRevResearch.4.013051}
  {\bibfield  {journal} {\bibinfo  {journal} {Phys. Rev. Res.}\ }\textbf
  {\bibinfo {volume} {4}},\ \bibinfo {pages} {013051} (\bibinfo {year}
  {2022})}\BibitemShut {NoStop}%
\bibitem [{\citenamefont {Quinn}\ \emph {et~al.}(2023)\citenamefont {Quinn},
  \citenamefont {Metzner}, \citenamefont {Muldoon}, \citenamefont {Moore},
  \citenamefont {Brudney}, \citenamefont {Das}, \citenamefont {Allcock},\ and\
  \citenamefont {Joglekar}}]{Quinn2023}%
  \BibitemOpen
  \bibfield  {author} {\bibinfo {author} {\bibfnamefont {A.}~\bibnamefont
  {Quinn}}, \bibinfo {author} {\bibfnamefont {J.}~\bibnamefont {Metzner}},
  \bibinfo {author} {\bibfnamefont {J.~E.}\ \bibnamefont {Muldoon}}, \bibinfo
  {author} {\bibfnamefont {I.~D.}\ \bibnamefont {Moore}}, \bibinfo {author}
  {\bibfnamefont {S.}~\bibnamefont {Brudney}}, \bibinfo {author} {\bibfnamefont
  {S.}~\bibnamefont {Das}}, \bibinfo {author} {\bibfnamefont {D.~T.~C.}\
  \bibnamefont {Allcock}}, \ and\ \bibinfo {author} {\bibfnamefont {Y.~N.}\
  \bibnamefont {Joglekar}},\ }\href@noop {} {\enquote {\bibinfo {title}
  {Observing super-quantum correlations across the exceptional point in a
  single, two-level trapped ion},}\ } (\bibinfo {year} {2023}),\ \Eprint
  {http://arxiv.org/abs/2304.12413} {arXiv:2304.12413 [quant-ph]} \BibitemShut
  {NoStop}%
\bibitem [{\citenamefont {Brody}\ and\ \citenamefont
  {Graefe}(2012)}]{Brody2012}%
  \BibitemOpen
  \bibfield  {author} {\bibinfo {author} {\bibfnamefont {D.~C.}\ \bibnamefont
  {Brody}}\ and\ \bibinfo {author} {\bibfnamefont {E.-M.}\ \bibnamefont
  {Graefe}},\ }\bibfield  {title} {\enquote {\bibinfo {title} {Mixed-state
  evolution in the presence of gain and loss},}\ }\href {\doibase
  10.1103/PhysRevLett.109.230405} {\bibfield  {journal} {\bibinfo  {journal}
  {Phys. Rev. Lett.}\ }\textbf {\bibinfo {volume} {109}},\ \bibinfo {pages}
  {230405} (\bibinfo {year} {2012})}\BibitemShut {NoStop}%
\bibitem [{\citenamefont {Varma}\ \emph {et~al.}(2023)\citenamefont {Varma},
  \citenamefont {Muldoon}, \citenamefont {Paul}, \citenamefont {Joglekar},\
  and\ \citenamefont {Das}}]{Varma2023}%
  \BibitemOpen
  \bibfield  {author} {\bibinfo {author} {\bibfnamefont {A.~V.}\ \bibnamefont
  {Varma}}, \bibinfo {author} {\bibfnamefont {J.~E.}\ \bibnamefont {Muldoon}},
  \bibinfo {author} {\bibfnamefont {S.}~\bibnamefont {Paul}}, \bibinfo {author}
  {\bibfnamefont {Y.~N.}\ \bibnamefont {Joglekar}}, \ and\ \bibinfo {author}
  {\bibfnamefont {S.}~\bibnamefont {Das}},\ }\bibfield  {title} {\enquote
  {\bibinfo {title} {Extreme violation of the leggett-garg inequality in
  nonunitary dynamics with complex energies},}\ }\href {\doibase
  10.1103/PhysRevA.108.032202} {\bibfield  {journal} {\bibinfo  {journal}
  {Phys. Rev. A}\ }\textbf {\bibinfo {volume} {108}},\ \bibinfo {pages}
  {032202} (\bibinfo {year} {2023})}\BibitemShut {NoStop}%
\bibitem [{\citenamefont {Bian}\ \emph {et~al.}(2020)\citenamefont {Bian},
  \citenamefont {Xiao}, \citenamefont {Wang}, \citenamefont {Onanga},
  \citenamefont {Ruzicka}, \citenamefont {Yi}, \citenamefont {Joglekar},\ and\
  \citenamefont {Xue}}]{Xue2020}%
  \BibitemOpen
  \bibfield  {author} {\bibinfo {author} {\bibfnamefont {Z.}~\bibnamefont
  {Bian}}, \bibinfo {author} {\bibfnamefont {L.}~\bibnamefont {Xiao}}, \bibinfo
  {author} {\bibfnamefont {K.}~\bibnamefont {Wang}}, \bibinfo {author}
  {\bibfnamefont {F.~A.}\ \bibnamefont {Onanga}}, \bibinfo {author}
  {\bibfnamefont {F.}~\bibnamefont {Ruzicka}}, \bibinfo {author} {\bibfnamefont
  {W.}~\bibnamefont {Yi}}, \bibinfo {author} {\bibfnamefont {Y.~N.}\
  \bibnamefont {Joglekar}}, \ and\ \bibinfo {author} {\bibfnamefont
  {P.}~\bibnamefont {Xue}},\ }\bibfield  {title} {\enquote {\bibinfo {title}
  {Quantum information dynamics in a high-dimensional parity-time-symmetric
  system},}\ }\href {\doibase 10.1103/PhysRevA.102.030201} {\bibfield
  {journal} {\bibinfo  {journal} {Phys. Rev. A}\ }\textbf {\bibinfo {volume}
  {102}},\ \bibinfo {pages} {030201} (\bibinfo {year} {2020})}\BibitemShut
  {NoStop}%
\bibitem [{\citenamefont {Reck}\ \emph {et~al.}(1994)\citenamefont {Reck},
  \citenamefont {Zeilinger}, \citenamefont {Bernstein},\ and\ \citenamefont
  {Bertani}}]{Reck1994Experimental}%
  \BibitemOpen
  \bibfield  {author} {\bibinfo {author} {\bibfnamefont {M.}~\bibnamefont
  {Reck}}, \bibinfo {author} {\bibfnamefont {A.}~\bibnamefont {Zeilinger}},
  \bibinfo {author} {\bibfnamefont {H.~J.}\ \bibnamefont {Bernstein}}, \ and\
  \bibinfo {author} {\bibfnamefont {P.}~\bibnamefont {Bertani}},\ }\bibfield
  {title} {\enquote {\bibinfo {title} {Experimental realization of any discrete
  unitary operator},}\ }\href {\doibase 10.1103/PhysRevLett.73.58} {\bibfield
  {journal} {\bibinfo  {journal} {Phys. Rev. Lett.}\ }\textbf {\bibinfo
  {volume} {73}},\ \bibinfo {pages} {58--61} (\bibinfo {year}
  {1994})}\BibitemShut {NoStop}%
\bibitem [{\citenamefont {Clements}\ \emph {et~al.}(2016)\citenamefont
  {Clements}, \citenamefont {Humphreys}, \citenamefont {Metcalf}, \citenamefont
  {Kolthammer},\ and\ \citenamefont {Walmsley}}]{Clements2016Optimal}%
  \BibitemOpen
  \bibfield  {author} {\bibinfo {author} {\bibfnamefont {W.~R.}\ \bibnamefont
  {Clements}}, \bibinfo {author} {\bibfnamefont {P.~C.}\ \bibnamefont
  {Humphreys}}, \bibinfo {author} {\bibfnamefont {B.~J.}\ \bibnamefont
  {Metcalf}}, \bibinfo {author} {\bibfnamefont {W.~S.}\ \bibnamefont
  {Kolthammer}}, \ and\ \bibinfo {author} {\bibfnamefont {I.~A.}\ \bibnamefont
  {Walmsley}},\ }\bibfield  {title} {\enquote {\bibinfo {title} {Optimal design
  for universal multiport interferometers},}\ }\href {\doibase
  10.1364/OPTICA.3.001460} {\bibfield  {journal} {\bibinfo  {journal} {Optica}\
  }\textbf {\bibinfo {volume} {3}},\ \bibinfo {pages} {1460--1465} (\bibinfo
  {year} {2016})}\BibitemShut {NoStop}%
\bibitem [{\citenamefont {Arvind}\ \emph {et~al.}(1995)\citenamefont {Arvind},
  \citenamefont {Dutta}, \citenamefont {Mukunda},\ and\ \citenamefont
  {Simon}}]{Arvind1995}%
  \BibitemOpen
  \bibfield  {author} {\bibinfo {author} {\bibnamefont {Arvind}}, \bibinfo
  {author} {\bibfnamefont {B.}~\bibnamefont {Dutta}}, \bibinfo {author}
  {\bibfnamefont {N.}~\bibnamefont {Mukunda}}, \ and\ \bibinfo {author}
  {\bibfnamefont {R.}~\bibnamefont {Simon}},\ }\bibfield  {title} {\enquote
  {\bibinfo {title} {The real symplectic groups in quantum mechanics and
  optics},}\ }\href {\doibase 10.1007/bf02848172} {\bibfield  {journal}
  {\bibinfo  {journal} {Pramana}\ }\textbf {\bibinfo {volume} {45}},\ \bibinfo
  {pages} {471--497} (\bibinfo {year} {1995})}\BibitemShut {NoStop}%
\bibitem [{\citenamefont {W\"{u}nsche}(2000)}]{Wnsche2000}%
  \BibitemOpen
  \bibfield  {author} {\bibinfo {author} {\bibfnamefont {A.}~\bibnamefont
  {W\"{u}nsche}},\ }\bibfield  {title} {\enquote {\bibinfo {title} {Symplectic
  groups in quantum optics},}\ }\href {\doibase 10.1088/1464-4266/2/2/302}
  {\bibfield  {journal} {\bibinfo  {journal} {Journal of Optics B: Quantum and
  Semiclassical Optics}\ }\textbf {\bibinfo {volume} {2}},\ \bibinfo {pages}
  {73--80} (\bibinfo {year} {2000})}\BibitemShut {NoStop}%
\bibitem [{\citenamefont {Braunstein}(2005)}]{Braunstein2005Squeezing}%
  \BibitemOpen
  \bibfield  {author} {\bibinfo {author} {\bibfnamefont {S.~L.}\ \bibnamefont
  {Braunstein}},\ }\bibfield  {title} {\enquote {\bibinfo {title} {Squeezing as
  an irreducible resource},}\ }\href {\doibase 10.1103/PhysRevA.71.055801}
  {\bibfield  {journal} {\bibinfo  {journal} {Phys. Rev. A}\ }\textbf {\bibinfo
  {volume} {71}},\ \bibinfo {pages} {055801} (\bibinfo {year}
  {2005})}\BibitemShut {NoStop}%
\bibitem [{\citenamefont {Joglekar}(2010)}]{Joglekar2010a}%
  \BibitemOpen
  \bibfield  {author} {\bibinfo {author} {\bibfnamefont {Y.~N.}\ \bibnamefont
  {Joglekar}},\ }\bibfield  {title} {\enquote {\bibinfo {title} {Mapping
  between hamiltonians with attractive and repulsive potentials on a
  lattice},}\ }\href {\doibase 10.1103/PhysRevA.82.044101} {\bibfield
  {journal} {\bibinfo  {journal} {Phys. Rev. A}\ }\textbf {\bibinfo {volume}
  {82}},\ \bibinfo {pages} {044101} (\bibinfo {year} {2010})}\BibitemShut
  {NoStop}%
\bibitem [{\citenamefont {Balian}\ and\ \citenamefont
  {Brezin}(1969)}]{Balian1969}%
  \BibitemOpen
  \bibfield  {author} {\bibinfo {author} {\bibfnamefont {R.}~\bibnamefont
  {Balian}}\ and\ \bibinfo {author} {\bibfnamefont {E.}~\bibnamefont
  {Brezin}},\ }\bibfield  {title} {\enquote {\bibinfo {title} {Nonunitary
  bogoliubov transformations and extension of wick's theorem},}\ }\href
  {\doibase 10.1007/bf02710281} {\bibfield  {journal} {\bibinfo  {journal} {Il
  Nuovo Cimento B Series 10}\ }\textbf {\bibinfo {volume} {64}},\ \bibinfo
  {pages} {37--55} (\bibinfo {year} {1969})}\BibitemShut {NoStop}%
\bibitem [{\citenamefont {Colpa}(1978)}]{Colpa1978}%
  \BibitemOpen
  \bibfield  {author} {\bibinfo {author} {\bibfnamefont {J.}~\bibnamefont
  {Colpa}},\ }\bibfield  {title} {\enquote {\bibinfo {title} {Diagonalization
  of the quadratic boson hamiltonian},}\ }\href {\doibase
  10.1016/0378-4371(78)90160-7} {\bibfield  {journal} {\bibinfo  {journal}
  {Physica A: Statistical Mechanics and its Applications}\ }\textbf {\bibinfo
  {volume} {93}},\ \bibinfo {pages} {327--353} (\bibinfo {year}
  {1978})}\BibitemShut {NoStop}%
\bibitem [{\citenamefont {Tsallis}(1978)}]{Tsallis1978}%
  \BibitemOpen
  \bibfield  {author} {\bibinfo {author} {\bibfnamefont {C.}~\bibnamefont
  {Tsallis}},\ }\bibfield  {title} {\enquote {\bibinfo {title} {Diagonalization
  methods for the general bilinear hamiltonian of an assembly of bosons},}\
  }\href {\doibase 10.1063/1.523549} {\bibfield  {journal} {\bibinfo  {journal}
  {Journal of Mathematical Physics}\ }\textbf {\bibinfo {volume} {19}},\
  \bibinfo {pages} {277--286} (\bibinfo {year} {1978})}\BibitemShut {NoStop}%
\bibitem [{\citenamefont {Broadbridge}\ and\ \citenamefont
  {Hurst}(1981)}]{Broadbridge1981}%
  \BibitemOpen
  \bibfield  {author} {\bibinfo {author} {\bibfnamefont {P.}~\bibnamefont
  {Broadbridge}}\ and\ \bibinfo {author} {\bibfnamefont {C.}~\bibnamefont
  {Hurst}},\ }\bibfield  {title} {\enquote {\bibinfo {title} {Canonical forms
  for quadratic hamiltonians},}\ }\href {\doibase 10.1016/0378-4371(81)90164-3}
  {\bibfield  {journal} {\bibinfo  {journal} {Physica A: Statistical Mechanics
  and its Applications}\ }\textbf {\bibinfo {volume} {108}},\ \bibinfo {pages}
  {39--62} (\bibinfo {year} {1981})}\BibitemShut {NoStop}%
\bibitem [{\citenamefont {Ruzicka}, \citenamefont {Agarwal},\ and\
  \citenamefont {Joglekar}(2021)}]{Ruzicka2021}%
  \BibitemOpen
  \bibfield  {author} {\bibinfo {author} {\bibfnamefont {F.}~\bibnamefont
  {Ruzicka}}, \bibinfo {author} {\bibfnamefont {K.~S.}\ \bibnamefont
  {Agarwal}}, \ and\ \bibinfo {author} {\bibfnamefont {Y.~N.}\ \bibnamefont
  {Joglekar}},\ }\bibfield  {title} {\enquote {\bibinfo {title} {Conserved
  quantities, exceptional points, and antilinear symmetries in non-hermitian
  systems},}\ }\href {\doibase 10.1088/1742-6596/2038/1/012021} {\bibfield
  {journal} {\bibinfo  {journal} {Journal of Physics: Conference Series}\
  }\textbf {\bibinfo {volume} {2038}},\ \bibinfo {pages} {012021} (\bibinfo
  {year} {2021})}\BibitemShut {NoStop}%
\bibitem [{\citenamefont {Buluta}\ and\ \citenamefont
  {Nori}(2009)}]{Buluta2009}%
  \BibitemOpen
  \bibfield  {author} {\bibinfo {author} {\bibfnamefont {I.}~\bibnamefont
  {Buluta}}\ and\ \bibinfo {author} {\bibfnamefont {F.}~\bibnamefont {Nori}},\
  }\bibfield  {title} {\enquote {\bibinfo {title} {Quantum simulators},}\
  }\href {\doibase 10.1126/science.1177838} {\bibfield  {journal} {\bibinfo
  {journal} {Science}\ }\textbf {\bibinfo {volume} {326}},\ \bibinfo {pages}
  {108--111} (\bibinfo {year} {2009})}\BibitemShut {NoStop}%
\bibitem [{\citenamefont {Altman}\ \emph {et~al.}(2021)\citenamefont {Altman},
  \citenamefont {Brown}, \citenamefont {Carleo}, \citenamefont {Carr},
  \citenamefont {Demler}, \citenamefont {Chin}, \citenamefont {DeMarco},
  \citenamefont {Economou}, \citenamefont {Eriksson}, \citenamefont {Fu},
  \citenamefont {Greiner}, \citenamefont {Hazzard}, \citenamefont {Hulet},
  \citenamefont {Koll\'ar}, \citenamefont {Lev}, \citenamefont {Lukin},
  \citenamefont {Ma}, \citenamefont {Mi}, \citenamefont {Misra}, \citenamefont
  {Monroe}, \citenamefont {Murch}, \citenamefont {Nazario}, \citenamefont {Ni},
  \citenamefont {Potter}, \citenamefont {Roushan}, \citenamefont {Saffman},
  \citenamefont {Schleier-Smith}, \citenamefont {Siddiqi}, \citenamefont
  {Simmonds}, \citenamefont {Singh}, \citenamefont {Spielman}, \citenamefont
  {Temme}, \citenamefont {Weiss}, \citenamefont {Vu\ifmmode \check{c}\else
  \v{c}\fi{}kovi\ifmmode~\acute{c}\else \'{c}\fi{}}, \citenamefont
  {Vuleti\ifmmode~\acute{c}\else \'{c}\fi{}}, \citenamefont {Ye},\ and\
  \citenamefont {Zwierlein}}]{Altman2021}%
  \BibitemOpen
  \bibfield  {author} {\bibinfo {author} {\bibfnamefont {E.}~\bibnamefont
  {Altman}}, \bibinfo {author} {\bibfnamefont {K.~R.}\ \bibnamefont {Brown}},
  \bibinfo {author} {\bibfnamefont {G.}~\bibnamefont {Carleo}}, \bibinfo
  {author} {\bibfnamefont {L.~D.}\ \bibnamefont {Carr}}, \bibinfo {author}
  {\bibfnamefont {E.}~\bibnamefont {Demler}}, \bibinfo {author} {\bibfnamefont
  {C.}~\bibnamefont {Chin}}, \bibinfo {author} {\bibfnamefont {B.}~\bibnamefont
  {DeMarco}}, \bibinfo {author} {\bibfnamefont {S.~E.}\ \bibnamefont
  {Economou}}, \bibinfo {author} {\bibfnamefont {M.~A.}\ \bibnamefont
  {Eriksson}}, \bibinfo {author} {\bibfnamefont {K.-M.~C.}\ \bibnamefont {Fu}},
  \bibinfo {author} {\bibfnamefont {M.}~\bibnamefont {Greiner}}, \bibinfo
  {author} {\bibfnamefont {K.~R.}\ \bibnamefont {Hazzard}}, \bibinfo {author}
  {\bibfnamefont {R.~G.}\ \bibnamefont {Hulet}}, \bibinfo {author}
  {\bibfnamefont {A.~J.}\ \bibnamefont {Koll\'ar}}, \bibinfo {author}
  {\bibfnamefont {B.~L.}\ \bibnamefont {Lev}}, \bibinfo {author} {\bibfnamefont
  {M.~D.}\ \bibnamefont {Lukin}}, \bibinfo {author} {\bibfnamefont
  {R.}~\bibnamefont {Ma}}, \bibinfo {author} {\bibfnamefont {X.}~\bibnamefont
  {Mi}}, \bibinfo {author} {\bibfnamefont {S.}~\bibnamefont {Misra}}, \bibinfo
  {author} {\bibfnamefont {C.}~\bibnamefont {Monroe}}, \bibinfo {author}
  {\bibfnamefont {K.}~\bibnamefont {Murch}}, \bibinfo {author} {\bibfnamefont
  {Z.}~\bibnamefont {Nazario}}, \bibinfo {author} {\bibfnamefont {K.-K.}\
  \bibnamefont {Ni}}, \bibinfo {author} {\bibfnamefont {A.~C.}\ \bibnamefont
  {Potter}}, \bibinfo {author} {\bibfnamefont {P.}~\bibnamefont {Roushan}},
  \bibinfo {author} {\bibfnamefont {M.}~\bibnamefont {Saffman}}, \bibinfo
  {author} {\bibfnamefont {M.}~\bibnamefont {Schleier-Smith}}, \bibinfo
  {author} {\bibfnamefont {I.}~\bibnamefont {Siddiqi}}, \bibinfo {author}
  {\bibfnamefont {R.}~\bibnamefont {Simmonds}}, \bibinfo {author}
  {\bibfnamefont {M.}~\bibnamefont {Singh}}, \bibinfo {author} {\bibfnamefont
  {I.}~\bibnamefont {Spielman}}, \bibinfo {author} {\bibfnamefont
  {K.}~\bibnamefont {Temme}}, \bibinfo {author} {\bibfnamefont {D.~S.}\
  \bibnamefont {Weiss}}, \bibinfo {author} {\bibfnamefont {J.}~\bibnamefont
  {Vu\ifmmode \check{c}\else \v{c}\fi{}kovi\ifmmode~\acute{c}\else
  \'{c}\fi{}}}, \bibinfo {author} {\bibfnamefont {V.}~\bibnamefont
  {Vuleti\ifmmode~\acute{c}\else \'{c}\fi{}}}, \bibinfo {author} {\bibfnamefont
  {J.}~\bibnamefont {Ye}}, \ and\ \bibinfo {author} {\bibfnamefont
  {M.}~\bibnamefont {Zwierlein}},\ }\bibfield  {title} {\enquote {\bibinfo
  {title} {Quantum simulators: Architectures and opportunities},}\ }\href
  {\doibase 10.1103/PRXQuantum.2.017003} {\bibfield  {journal} {\bibinfo
  {journal} {PRX Quantum}\ }\textbf {\bibinfo {volume} {2}},\ \bibinfo {pages}
  {017003} (\bibinfo {year} {2021})}\BibitemShut {NoStop}%
\bibitem [{\citenamefont {Daley}\ \emph {et~al.}(2022)\citenamefont {Daley},
  \citenamefont {Bloch}, \citenamefont {Kokail}, \citenamefont {Flannigan},
  \citenamefont {Pearson}, \citenamefont {Troyer},\ and\ \citenamefont
  {Zoller}}]{Daley2022}%
  \BibitemOpen
  \bibfield  {author} {\bibinfo {author} {\bibfnamefont {A.~J.}\ \bibnamefont
  {Daley}}, \bibinfo {author} {\bibfnamefont {I.}~\bibnamefont {Bloch}},
  \bibinfo {author} {\bibfnamefont {C.}~\bibnamefont {Kokail}}, \bibinfo
  {author} {\bibfnamefont {S.}~\bibnamefont {Flannigan}}, \bibinfo {author}
  {\bibfnamefont {N.}~\bibnamefont {Pearson}}, \bibinfo {author} {\bibfnamefont
  {M.}~\bibnamefont {Troyer}}, \ and\ \bibinfo {author} {\bibfnamefont
  {P.}~\bibnamefont {Zoller}},\ }\bibfield  {title} {\enquote {\bibinfo {title}
  {Practical quantum advantage in quantum simulation},}\ }\href {\doibase
  10.1038/s41586-022-04940-6} {\bibfield  {journal} {\bibinfo  {journal}
  {Nature}\ }\textbf {\bibinfo {volume} {607}},\ \bibinfo {pages} {667--676}
  (\bibinfo {year} {2022})}\BibitemShut {NoStop}%
\bibitem [{\citenamefont {Harrington}, \citenamefont {Mueller},\ and\
  \citenamefont {Murch}(2022)}]{Harrington2022}%
  \BibitemOpen
  \bibfield  {author} {\bibinfo {author} {\bibfnamefont {P.~M.}\ \bibnamefont
  {Harrington}}, \bibinfo {author} {\bibfnamefont {E.~J.}\ \bibnamefont
  {Mueller}}, \ and\ \bibinfo {author} {\bibfnamefont {K.~W.}\ \bibnamefont
  {Murch}},\ }\bibfield  {title} {\enquote {\bibinfo {title} {Engineered
  dissipation for quantum information science},}\ }\href {\doibase
  10.1038/s42254-022-00494-8} {\bibfield  {journal} {\bibinfo  {journal}
  {Nature Reviews Physics}\ }\textbf {\bibinfo {volume} {4}},\ \bibinfo {pages}
  {660--671} (\bibinfo {year} {2022})}\BibitemShut {NoStop}%
\bibitem [{\citenamefont {Knill}, \citenamefont {Laflamme},\ and\ \citenamefont
  {Milburn}(2001)}]{Knill2001}%
  \BibitemOpen
  \bibfield  {author} {\bibinfo {author} {\bibfnamefont {E.}~\bibnamefont
  {Knill}}, \bibinfo {author} {\bibfnamefont {R.}~\bibnamefont {Laflamme}}, \
  and\ \bibinfo {author} {\bibfnamefont {G.~J.}\ \bibnamefont {Milburn}},\
  }\bibfield  {title} {\enquote {\bibinfo {title} {A scheme for efficient
  quantum computation with linear optics},}\ }\href {\doibase 10.1038/35051009}
  {\bibfield  {journal} {\bibinfo  {journal} {Nature}\ }\textbf {\bibinfo
  {volume} {409}},\ \bibinfo {pages} {46--52} (\bibinfo {year}
  {2001})}\BibitemShut {NoStop}%
\bibitem [{\citenamefont {Carolan}\ \emph {et~al.}(2015)\citenamefont
  {Carolan}, \citenamefont {Harrold}, \citenamefont {Sparrow}, \citenamefont
  {Mart{\'{\i}}n-L{\'{o}}pez}, \citenamefont {Russell}, \citenamefont
  {Silverstone}, \citenamefont {Shadbolt}, \citenamefont {Matsuda},
  \citenamefont {Oguma}, \citenamefont {Itoh}, \citenamefont {Marshall},
  \citenamefont {Thompson}, \citenamefont {Matthews}, \citenamefont
  {Hashimoto}, \citenamefont {O'Brien},\ and\ \citenamefont
  {Laing}}]{Carolan2015}%
  \BibitemOpen
  \bibfield  {author} {\bibinfo {author} {\bibfnamefont {J.}~\bibnamefont
  {Carolan}}, \bibinfo {author} {\bibfnamefont {C.}~\bibnamefont {Harrold}},
  \bibinfo {author} {\bibfnamefont {C.}~\bibnamefont {Sparrow}}, \bibinfo
  {author} {\bibfnamefont {E.}~\bibnamefont {Mart{\'{\i}}n-L{\'{o}}pez}},
  \bibinfo {author} {\bibfnamefont {N.~J.}\ \bibnamefont {Russell}}, \bibinfo
  {author} {\bibfnamefont {J.~W.}\ \bibnamefont {Silverstone}}, \bibinfo
  {author} {\bibfnamefont {P.~J.}\ \bibnamefont {Shadbolt}}, \bibinfo {author}
  {\bibfnamefont {N.}~\bibnamefont {Matsuda}}, \bibinfo {author} {\bibfnamefont
  {M.}~\bibnamefont {Oguma}}, \bibinfo {author} {\bibfnamefont
  {M.}~\bibnamefont {Itoh}}, \bibinfo {author} {\bibfnamefont {G.~D.}\
  \bibnamefont {Marshall}}, \bibinfo {author} {\bibfnamefont {M.~G.}\
  \bibnamefont {Thompson}}, \bibinfo {author} {\bibfnamefont {J.~C.~F.}\
  \bibnamefont {Matthews}}, \bibinfo {author} {\bibfnamefont {T.}~\bibnamefont
  {Hashimoto}}, \bibinfo {author} {\bibfnamefont {J.~L.}\ \bibnamefont
  {O'Brien}}, \ and\ \bibinfo {author} {\bibfnamefont {A.}~\bibnamefont
  {Laing}},\ }\bibfield  {title} {\enquote {\bibinfo {title} {Universal linear
  optics},}\ }\href {\doibase 10.1126/science.aab3642} {\bibfield  {journal}
  {\bibinfo  {journal} {Science}\ }\textbf {\bibinfo {volume} {349}},\ \bibinfo
  {pages} {711--716} (\bibinfo {year} {2015})}\BibitemShut {NoStop}%
\bibitem [{\citenamefont {Sparrow}\ \emph {et~al.}(2018)\citenamefont
  {Sparrow}, \citenamefont {Mart{\'\i}n-L{\'o}pez}, \citenamefont {Maraviglia},
  \citenamefont {Neville}, \citenamefont {Harrold}, \citenamefont {Carolan},
  \citenamefont {Joglekar}, \citenamefont {Hashimoto}, \citenamefont {Matsuda},
  \citenamefont {O'Brien}, \citenamefont {Tew},\ and\ \citenamefont
  {Laing}}]{Sparrow2018-Nature}%
  \BibitemOpen
  \bibfield  {author} {\bibinfo {author} {\bibfnamefont {C.}~\bibnamefont
  {Sparrow}}, \bibinfo {author} {\bibfnamefont {E.}~\bibnamefont
  {Mart{\'\i}n-L{\'o}pez}}, \bibinfo {author} {\bibfnamefont {N.}~\bibnamefont
  {Maraviglia}}, \bibinfo {author} {\bibfnamefont {A.}~\bibnamefont {Neville}},
  \bibinfo {author} {\bibfnamefont {C.}~\bibnamefont {Harrold}}, \bibinfo
  {author} {\bibfnamefont {J.}~\bibnamefont {Carolan}}, \bibinfo {author}
  {\bibfnamefont {Y.~N.}\ \bibnamefont {Joglekar}}, \bibinfo {author}
  {\bibfnamefont {T.}~\bibnamefont {Hashimoto}}, \bibinfo {author}
  {\bibfnamefont {N.}~\bibnamefont {Matsuda}}, \bibinfo {author} {\bibfnamefont
  {J.~L.}\ \bibnamefont {O'Brien}}, \bibinfo {author} {\bibfnamefont {D.~P.}\
  \bibnamefont {Tew}}, \ and\ \bibinfo {author} {\bibfnamefont
  {A.}~\bibnamefont {Laing}},\ }\bibfield  {title} {\enquote {\bibinfo {title}
  {Simulating the vibrational quantum dynamics of molecules using photonics},}\
  }\href {\doibase 10.1038/s41586-018-0152-9} {\bibfield  {journal} {\bibinfo
  {journal} {Nature}\ }\textbf {\bibinfo {volume} {557}},\ \bibinfo {pages}
  {660--667} (\bibinfo {year} {2018})}\BibitemShut {NoStop}%
\bibitem [{\citenamefont {Tillmann}\ \emph {et~al.}(2013)\citenamefont
  {Tillmann}, \citenamefont {Daki{\'{c}}}, \citenamefont {Heilmann},
  \citenamefont {Nolte}, \citenamefont {Szameit},\ and\ \citenamefont
  {Walther}}]{Tillmann2013}%
  \BibitemOpen
  \bibfield  {author} {\bibinfo {author} {\bibfnamefont {M.}~\bibnamefont
  {Tillmann}}, \bibinfo {author} {\bibfnamefont {B.}~\bibnamefont
  {Daki{\'{c}}}}, \bibinfo {author} {\bibfnamefont {R.}~\bibnamefont
  {Heilmann}}, \bibinfo {author} {\bibfnamefont {S.}~\bibnamefont {Nolte}},
  \bibinfo {author} {\bibfnamefont {A.}~\bibnamefont {Szameit}}, \ and\
  \bibinfo {author} {\bibfnamefont {P.}~\bibnamefont {Walther}},\ }\bibfield
  {title} {\enquote {\bibinfo {title} {Experimental boson sampling},}\ }\href
  {\doibase 10.1038/nphoton.2013.102} {\bibfield  {journal} {\bibinfo
  {journal} {Nature Photonics}\ }\textbf {\bibinfo {volume} {7}},\ \bibinfo
  {pages} {540--544} (\bibinfo {year} {2013})}\BibitemShut {NoStop}%
\bibitem [{\citenamefont {Spagnolo}\ \emph {et~al.}(2014)\citenamefont
  {Spagnolo}, \citenamefont {Vitelli}, \citenamefont {Bentivegna},
  \citenamefont {Brod}, \citenamefont {Crespi}, \citenamefont {Flamini},
  \citenamefont {Giacomini}, \citenamefont {Milani}, \citenamefont {Ramponi},
  \citenamefont {Mataloni}, \citenamefont {Osellame}, \citenamefont
  {Galv{\~{a}}o},\ and\ \citenamefont {Sciarrino}}]{Spagnolo2014}%
  \BibitemOpen
  \bibfield  {author} {\bibinfo {author} {\bibfnamefont {N.}~\bibnamefont
  {Spagnolo}}, \bibinfo {author} {\bibfnamefont {C.}~\bibnamefont {Vitelli}},
  \bibinfo {author} {\bibfnamefont {M.}~\bibnamefont {Bentivegna}}, \bibinfo
  {author} {\bibfnamefont {D.~J.}\ \bibnamefont {Brod}}, \bibinfo {author}
  {\bibfnamefont {A.}~\bibnamefont {Crespi}}, \bibinfo {author} {\bibfnamefont
  {F.}~\bibnamefont {Flamini}}, \bibinfo {author} {\bibfnamefont
  {S.}~\bibnamefont {Giacomini}}, \bibinfo {author} {\bibfnamefont
  {G.}~\bibnamefont {Milani}}, \bibinfo {author} {\bibfnamefont
  {R.}~\bibnamefont {Ramponi}}, \bibinfo {author} {\bibfnamefont
  {P.}~\bibnamefont {Mataloni}}, \bibinfo {author} {\bibfnamefont
  {R.}~\bibnamefont {Osellame}}, \bibinfo {author} {\bibfnamefont {E.~F.}\
  \bibnamefont {Galv{\~{a}}o}}, \ and\ \bibinfo {author} {\bibfnamefont
  {F.}~\bibnamefont {Sciarrino}},\ }\bibfield  {title} {\enquote {\bibinfo
  {title} {Experimental validation of photonic boson sampling},}\ }\href
  {\doibase 10.1038/nphoton.2014.135} {\bibfield  {journal} {\bibinfo
  {journal} {Nature Photonics}\ }\textbf {\bibinfo {volume} {8}},\ \bibinfo
  {pages} {615--620} (\bibinfo {year} {2014})}\BibitemShut {NoStop}%
\bibitem [{\citenamefont {Carolan}\ \emph {et~al.}(2014)\citenamefont
  {Carolan}, \citenamefont {Meinecke}, \citenamefont {Shadbolt}, \citenamefont
  {Russell}, \citenamefont {Ismail}, \citenamefont {W\"{o}rhoff}, \citenamefont
  {Rudolph}, \citenamefont {Thompson}, \citenamefont
  {O{\textquotesingle}Brien}, \citenamefont {Matthews},\ and\ \citenamefont
  {Laing}}]{Carolan2014}%
  \BibitemOpen
  \bibfield  {author} {\bibinfo {author} {\bibfnamefont {J.}~\bibnamefont
  {Carolan}}, \bibinfo {author} {\bibfnamefont {J.~D.~A.}\ \bibnamefont
  {Meinecke}}, \bibinfo {author} {\bibfnamefont {P.~J.}\ \bibnamefont
  {Shadbolt}}, \bibinfo {author} {\bibfnamefont {N.~J.}\ \bibnamefont
  {Russell}}, \bibinfo {author} {\bibfnamefont {N.}~\bibnamefont {Ismail}},
  \bibinfo {author} {\bibfnamefont {K.}~\bibnamefont {W\"{o}rhoff}}, \bibinfo
  {author} {\bibfnamefont {T.}~\bibnamefont {Rudolph}}, \bibinfo {author}
  {\bibfnamefont {M.~G.}\ \bibnamefont {Thompson}}, \bibinfo {author}
  {\bibfnamefont {J.~L.}\ \bibnamefont {O{\textquotesingle}Brien}}, \bibinfo
  {author} {\bibfnamefont {J.~C.~F.}\ \bibnamefont {Matthews}}, \ and\ \bibinfo
  {author} {\bibfnamefont {A.}~\bibnamefont {Laing}},\ }\bibfield  {title}
  {\enquote {\bibinfo {title} {On the experimental verification of quantum
  complexity in linear optics},}\ }\href {\doibase 10.1038/nphoton.2014.152}
  {\bibfield  {journal} {\bibinfo  {journal} {Nature Photonics}\ }\textbf
  {\bibinfo {volume} {8}},\ \bibinfo {pages} {621--626} (\bibinfo {year}
  {2014})}\BibitemShut {NoStop}%
\bibitem [{\citenamefont {Horn}\ and\ \citenamefont
  {Johnson}(1985)}]{Horn1985}%
  \BibitemOpen
  \bibfield  {author} {\bibinfo {author} {\bibfnamefont {R.~A.}\ \bibnamefont
  {Horn}}\ and\ \bibinfo {author} {\bibfnamefont {C.~R.}\ \bibnamefont
  {Johnson}},\ }\href {\doibase 10.1017/cbo9780511810817} {\emph {\bibinfo
  {title} {Matrix Analysis}}}\ (\bibinfo  {publisher} {Cambridge University
  Press},\ \bibinfo {year} {1985})\BibitemShut {NoStop}%
\bibitem [{\citenamefont {Tischler}, \citenamefont {Rockstuhl},\ and\
  \citenamefont {S\l{}owik}(2018)}]{Tischler2018Arbitrary}%
  \BibitemOpen
  \bibfield  {author} {\bibinfo {author} {\bibfnamefont {N.}~\bibnamefont
  {Tischler}}, \bibinfo {author} {\bibfnamefont {C.}~\bibnamefont {Rockstuhl}},
  \ and\ \bibinfo {author} {\bibfnamefont {K.}~\bibnamefont {S\l{}owik}},\
  }\bibfield  {title} {\enquote {\bibinfo {title} {Quantum optical realization
  of arbitrary linear transformations allowing for loss and gain},}\ }\href
  {\doibase 10.1103/PhysRevX.8.021017} {\bibfield  {journal} {\bibinfo
  {journal} {Phys. Rev. X}\ }\textbf {\bibinfo {volume} {8}},\ \bibinfo {pages}
  {021017} (\bibinfo {year} {2018})}\BibitemShut {NoStop}%
\bibitem [{\citenamefont {Barthel}\ and\ \citenamefont
  {Zhang}(2020)}]{Barthel2020}%
  \BibitemOpen
  \bibfield  {author} {\bibinfo {author} {\bibfnamefont {T.}~\bibnamefont
  {Barthel}}\ and\ \bibinfo {author} {\bibfnamefont {Y.}~\bibnamefont
  {Zhang}},\ }\bibfield  {title} {\enquote {\bibinfo {title} {Optimized
  lie{\textendash}trotter{\textendash}suzuki decompositions for two and three
  non-commuting terms},}\ }\href {\doibase 10.1016/j.aop.2020.168165}
  {\bibfield  {journal} {\bibinfo  {journal} {Annals of Physics}\ }\textbf
  {\bibinfo {volume} {418}},\ \bibinfo {pages} {168165} (\bibinfo {year}
  {2020})}\BibitemShut {NoStop}%
\bibitem [{\citenamefont {Clinton}, \citenamefont {Bausch},\ and\ \citenamefont
  {Cubitt}(2021)}]{Clinton2021}%
  \BibitemOpen
  \bibfield  {author} {\bibinfo {author} {\bibfnamefont {L.}~\bibnamefont
  {Clinton}}, \bibinfo {author} {\bibfnamefont {J.}~\bibnamefont {Bausch}}, \
  and\ \bibinfo {author} {\bibfnamefont {T.}~\bibnamefont {Cubitt}},\
  }\bibfield  {title} {\enquote {\bibinfo {title} {Hamiltonian simulation
  algorithms for near-term quantum hardware},}\ }\href {\doibase
  10.1038/s41467-021-25196-0} {\bibfield  {journal} {\bibinfo  {journal}
  {Nature Communications}\ }\textbf {\bibinfo {volume} {12}} (\bibinfo {year}
  {2021}),\ 10.1038/s41467-021-25196-0}\BibitemShut {NoStop}%
\bibitem [{\citenamefont {Childs}\ \emph {et~al.}(2021)\citenamefont {Childs},
  \citenamefont {Su}, \citenamefont {Tran}, \citenamefont {Wiebe},\ and\
  \citenamefont {Zhu}}]{Childs2021}%
  \BibitemOpen
  \bibfield  {author} {\bibinfo {author} {\bibfnamefont {A.~M.}\ \bibnamefont
  {Childs}}, \bibinfo {author} {\bibfnamefont {Y.}~\bibnamefont {Su}}, \bibinfo
  {author} {\bibfnamefont {M.~C.}\ \bibnamefont {Tran}}, \bibinfo {author}
  {\bibfnamefont {N.}~\bibnamefont {Wiebe}}, \ and\ \bibinfo {author}
  {\bibfnamefont {S.}~\bibnamefont {Zhu}},\ }\bibfield  {title} {\enquote
  {\bibinfo {title} {Theory of trotter error with commutator scaling},}\ }\href
  {\doibase 10.1103/PhysRevX.11.011020} {\bibfield  {journal} {\bibinfo
  {journal} {Phys. Rev. X}\ }\textbf {\bibinfo {volume} {11}},\ \bibinfo
  {pages} {011020} (\bibinfo {year} {2021})}\BibitemShut {NoStop}%
\bibitem [{\citenamefont {Antonosyan}, \citenamefont {Solntsev},\ and\
  \citenamefont {Sukhorukov}(2015)}]{Antonosyan2015}%
  \BibitemOpen
  \bibfield  {author} {\bibinfo {author} {\bibfnamefont {D.~A.}\ \bibnamefont
  {Antonosyan}}, \bibinfo {author} {\bibfnamefont {A.~S.}\ \bibnamefont
  {Solntsev}}, \ and\ \bibinfo {author} {\bibfnamefont {A.~A.}\ \bibnamefont
  {Sukhorukov}},\ }\bibfield  {title} {\enquote {\bibinfo {title} {Parity-time
  anti-symmetric parametric amplifier},}\ }\href {\doibase
  10.1364/ol.40.004575} {\bibfield  {journal} {\bibinfo  {journal} {Optics
  Letters}\ }\textbf {\bibinfo {volume} {40}},\ \bibinfo {pages} {4575}
  (\bibinfo {year} {2015})}\BibitemShut {NoStop}%
\bibitem [{\citenamefont {Suchkov}\ \emph {et~al.}(2016)\citenamefont
  {Suchkov}, \citenamefont {Sukhorukov}, \citenamefont {Huang}, \citenamefont
  {Dmitriev}, \citenamefont {Lee},\ and\ \citenamefont
  {Kivshar}}]{Suchkov2016}%
  \BibitemOpen
  \bibfield  {author} {\bibinfo {author} {\bibfnamefont {S.~V.}\ \bibnamefont
  {Suchkov}}, \bibinfo {author} {\bibfnamefont {A.~A.}\ \bibnamefont
  {Sukhorukov}}, \bibinfo {author} {\bibfnamefont {J.}~\bibnamefont {Huang}},
  \bibinfo {author} {\bibfnamefont {S.~V.}\ \bibnamefont {Dmitriev}}, \bibinfo
  {author} {\bibfnamefont {C.}~\bibnamefont {Lee}}, \ and\ \bibinfo {author}
  {\bibfnamefont {Y.~S.}\ \bibnamefont {Kivshar}},\ }\bibfield  {title}
  {\enquote {\bibinfo {title} {Nonlinear switching and solitons in
  {PT}-symmetric photonic systems},}\ }\href {\doibase 10.1002/lpor.201500227}
  {\bibfield  {journal} {\bibinfo  {journal} {Laser {\&} Photonics Reviews}\
  }\textbf {\bibinfo {volume} {10}},\ \bibinfo {pages} {177--213} (\bibinfo
  {year} {2016})}\BibitemShut {NoStop}%
\bibitem [{\citenamefont {Miri}\ and\ \citenamefont
  {Al{\`{u}}}(2016)}]{Miri2016}%
  \BibitemOpen
  \bibfield  {author} {\bibinfo {author} {\bibfnamefont {M.-A.}\ \bibnamefont
  {Miri}}\ and\ \bibinfo {author} {\bibfnamefont {A.}~\bibnamefont
  {Al{\`{u}}}},\ }\bibfield  {title} {\enquote {\bibinfo {title}
  {Nonlinearity-induced {PT}-symmetry without material gain},}\ }\href
  {\doibase 10.1088/1367-2630/18/6/065001} {\bibfield  {journal} {\bibinfo
  {journal} {New Journal of Physics}\ }\textbf {\bibinfo {volume} {18}},\
  \bibinfo {pages} {065001} (\bibinfo {year} {2016})}\BibitemShut {NoStop}%
\bibitem [{\citenamefont {Longhi}(2018)}]{Longhi2018}%
  \BibitemOpen
  \bibfield  {author} {\bibinfo {author} {\bibfnamefont {S.}~\bibnamefont
  {Longhi}},\ }\bibfield  {title} {\enquote {\bibinfo {title} {{PT} symmetry
  and antisymmetry by anti-hermitian wave coupling and nonlinear optical
  interactions},}\ }\href {\doibase 10.1364/ol.43.004025} {\bibfield  {journal}
  {\bibinfo  {journal} {Optics Letters}\ }\textbf {\bibinfo {volume} {43}},\
  \bibinfo {pages} {4025} (\bibinfo {year} {2018})}\BibitemShut {NoStop}%
\bibitem [{\citenamefont {Humire}\ \emph {et~al.}(2023)\citenamefont {Humire},
  \citenamefont {Z{\'{a}}rate}, \citenamefont {Joglekar},\ and\ \citenamefont
  {Garc{\'{\i}}a-{\~{N}}ustes}}]{Humire2023}%
  \BibitemOpen
  \bibfield  {author} {\bibinfo {author} {\bibfnamefont {F.~R.}\ \bibnamefont
  {Humire}}, \bibinfo {author} {\bibfnamefont {Y.~D.}\ \bibnamefont
  {Z{\'{a}}rate}}, \bibinfo {author} {\bibfnamefont {Y.~N.}\ \bibnamefont
  {Joglekar}}, \ and\ \bibinfo {author} {\bibfnamefont {M.~A.}\ \bibnamefont
  {Garc{\'{\i}}a-{\~{N}}ustes}},\ }\bibfield  {title} {\enquote {\bibinfo
  {title} {Classical rabi oscillations induced by unbalanced dissipation on a
  nonlinear dimer},}\ }\href {\doibase 10.1016/j.chaos.2023.113435} {\bibfield
  {journal} {\bibinfo  {journal} {Chaos, Solitons {\&} Fractals}\ }\textbf
  {\bibinfo {volume} {171}},\ \bibinfo {pages} {113435} (\bibinfo {year}
  {2023})}\BibitemShut {NoStop}%
\bibitem [{\citenamefont {Jiang}\ \emph {et~al.}(2019)\citenamefont {Jiang},
  \citenamefont {Mei}, \citenamefont {Zuo}, \citenamefont {Zhai}, \citenamefont
  {Li}, \citenamefont {Wen},\ and\ \citenamefont {Du}}]{Jiang2019}%
  \BibitemOpen
  \bibfield  {author} {\bibinfo {author} {\bibfnamefont {Y.}~\bibnamefont
  {Jiang}}, \bibinfo {author} {\bibfnamefont {Y.}~\bibnamefont {Mei}}, \bibinfo
  {author} {\bibfnamefont {Y.}~\bibnamefont {Zuo}}, \bibinfo {author}
  {\bibfnamefont {Y.}~\bibnamefont {Zhai}}, \bibinfo {author} {\bibfnamefont
  {J.}~\bibnamefont {Li}}, \bibinfo {author} {\bibfnamefont {J.}~\bibnamefont
  {Wen}}, \ and\ \bibinfo {author} {\bibfnamefont {S.}~\bibnamefont {Du}},\
  }\bibfield  {title} {\enquote {\bibinfo {title} {Anti-parity-time symmetric
  optical four-wave mixing in cold atoms},}\ }\href {\doibase
  10.1103/PhysRevLett.123.193604} {\bibfield  {journal} {\bibinfo  {journal}
  {Phys. Rev. Lett.}\ }\textbf {\bibinfo {volume} {123}},\ \bibinfo {pages}
  {193604} (\bibinfo {year} {2019})}\BibitemShut {NoStop}%
\bibitem [{\citenamefont {Wang}\ and\ \citenamefont
  {Clerk}(2019)}]{AshClerk2019}%
  \BibitemOpen
  \bibfield  {author} {\bibinfo {author} {\bibfnamefont {Y.-X.}\ \bibnamefont
  {Wang}}\ and\ \bibinfo {author} {\bibfnamefont {A.~A.}\ \bibnamefont
  {Clerk}},\ }\bibfield  {title} {\enquote {\bibinfo {title} {Non-hermitian
  dynamics without dissipation in quantum systems},}\ }\href {\doibase
  10.1103/PhysRevA.99.063834} {\bibfield  {journal} {\bibinfo  {journal} {Phys.
  Rev. A}\ }\textbf {\bibinfo {volume} {99}},\ \bibinfo {pages} {063834}
  (\bibinfo {year} {2019})}\BibitemShut {NoStop}%
\bibitem [{\citenamefont {Luo}, \citenamefont {Zhang},\ and\ \citenamefont
  {Du}(2022)}]{Luo2022}%
  \BibitemOpen
  \bibfield  {author} {\bibinfo {author} {\bibfnamefont {X.-W.}\ \bibnamefont
  {Luo}}, \bibinfo {author} {\bibfnamefont {C.}~\bibnamefont {Zhang}}, \ and\
  \bibinfo {author} {\bibfnamefont {S.}~\bibnamefont {Du}},\ }\bibfield
  {title} {\enquote {\bibinfo {title} {Quantum squeezing and sensing with
  pseudo-anti-parity-time symmetry},}\ }\href {\doibase
  10.1103/PhysRevLett.128.173602} {\bibfield  {journal} {\bibinfo  {journal}
  {Phys. Rev. Lett.}\ }\textbf {\bibinfo {volume} {128}},\ \bibinfo {pages}
  {173602} (\bibinfo {year} {2022})}\BibitemShut {NoStop}%
\bibitem [{\citenamefont {Gaikwad}\ \emph {et~al.}(2023)\citenamefont
  {Gaikwad}, \citenamefont {Kowsari}, \citenamefont {Chen},\ and\ \citenamefont
  {Murch}}]{Gaikwad2023}%
  \BibitemOpen
  \bibfield  {author} {\bibinfo {author} {\bibfnamefont {C.}~\bibnamefont
  {Gaikwad}}, \bibinfo {author} {\bibfnamefont {D.}~\bibnamefont {Kowsari}},
  \bibinfo {author} {\bibfnamefont {W.}~\bibnamefont {Chen}}, \ and\ \bibinfo
  {author} {\bibfnamefont {K.~W.}\ \bibnamefont {Murch}},\ }\href@noop {}
  {\enquote {\bibinfo {title} {Observing parity time symmetry breaking in a
  josephson parametric amplifier},}\ } (\bibinfo {year} {2023}),\ \Eprint
  {http://arxiv.org/abs/2306.14980} {arXiv:2306.14980 [quant-ph]} \BibitemShut
  {NoStop}%
\bibitem [{\citenamefont {Burd}\ \emph {et~al.}(2021)\citenamefont {Burd},
  \citenamefont {Srinivas}, \citenamefont {Knaack}, \citenamefont {Ge},
  \citenamefont {Wilson}, \citenamefont {Wineland}, \citenamefont {Leibfried},
  \citenamefont {Bollinger}, \citenamefont {Allcock},\ and\ \citenamefont
  {Slichter}}]{Burd2021}%
  \BibitemOpen
  \bibfield  {author} {\bibinfo {author} {\bibfnamefont {S.~C.}\ \bibnamefont
  {Burd}}, \bibinfo {author} {\bibfnamefont {R.}~\bibnamefont {Srinivas}},
  \bibinfo {author} {\bibfnamefont {H.~M.}\ \bibnamefont {Knaack}}, \bibinfo
  {author} {\bibfnamefont {W.}~\bibnamefont {Ge}}, \bibinfo {author}
  {\bibfnamefont {A.~C.}\ \bibnamefont {Wilson}}, \bibinfo {author}
  {\bibfnamefont {D.~J.}\ \bibnamefont {Wineland}}, \bibinfo {author}
  {\bibfnamefont {D.}~\bibnamefont {Leibfried}}, \bibinfo {author}
  {\bibfnamefont {J.~J.}\ \bibnamefont {Bollinger}}, \bibinfo {author}
  {\bibfnamefont {D.~T.~C.}\ \bibnamefont {Allcock}}, \ and\ \bibinfo {author}
  {\bibfnamefont {D.~H.}\ \bibnamefont {Slichter}},\ }\bibfield  {title}
  {\enquote {\bibinfo {title} {Quantum amplification of boson-mediated
  interactions},}\ }\href {\doibase 10.1038/s41567-021-01237-9} {\bibfield
  {journal} {\bibinfo  {journal} {Nature Physics}\ }\textbf {\bibinfo {volume}
  {17}},\ \bibinfo {pages} {898--902} (\bibinfo {year} {2021})}\BibitemShut
  {NoStop}%
\end{thebibliography}%


\begin{acknowledgments}
This work was supported by Bristol Benjamin Meaker Distinguished Visiting Professorship, ONR Grant No. N00014-21-1-2630, and EPSRC grant EP/LO15730/1. 
\end{acknowledgments}


\end{document}